\documentclass[10pt]{article}

\usepackage{amssymb, amsmath,amsthm, mathrsfs, bbm, MnSymbol, url}
\usepackage[dvips]{color}
\usepackage{svg}
\usepackage{dutchcal}

\usepackage{graphicx}
\usepackage{epstopdf}
\usepackage{mhchem}
\usepackage{wrapfig}

\newtheorem*{conclusion}{Conclusion}

\theoremstyle{remark}

\newenvironment{keywords}{\par}{\par}

\begin{document}

\title{Determination of output composition in reaction-advection-diffusion systems on network reactors}

\author{R. Feres\footnote{Dept. of Mathematics, Washington University,
Campus Box 1146, St. Louis, MO 63130},{\ }  E. Pasewark\footnotemark[1],{\ } G. Yablonsky\footnote{Department of Energy, Environmental  \& Chemical Engineering, McKelvey School of Engineering, Washington University, St. Louis, MO 63130. }}


\maketitle

\begin{abstract}{\small
We consider reaction-transport processes  in open reactors in which systems of first order reactions involving  a number of gas species and solid catalysts can occur at localized active regions. Reaction products flow out of the reactor into vacuum  conditions and are collected at an exit boundary. The {\em output composition problem} (OCP) is to determine the composition (molar fractions) of the collected gas after the reactor is fully emptied. We provide a solution to this problem in the form of a boundary-value problem for a system of time-independent partial differential equations. We then consider {\em network-like reactors}, which can be approximated by a network consisting of a collection of  nodes and $1$-dimensional branches, with reactions taking place at   nodes. For these, it is possible to solve the OCP in a simple and effective way, giving explicit formulas for the output composition as a function of the reaction coefficients and parameters  associated with the geometric configuration of the system. Several examples are given to illustrate the method.
 }\end{abstract}


\section{Introduction}\label{introduction}

\subsection{Physico-chemical motivation}
Reaction-transport problems, in particular reaction-diffusion problems, are among the most topical and widely studied   in chemical engineering. They have been posed and investigated  in classical works of chemical engineering since the very beginning of this discipline. See \cite{Aris,FK}.
In approaching such problems, different elements should be considered:
\begin{itemize}
\item The chemical reaction is complex, i.e., it consists of a set of reaction steps involving a reactive mixture.
\item Chemical reactions are always accompanied by transport.
\item Two types of transport are typically considered: {``forced''} propagation (advection) and/or {``self-propagation''} (diffusion).
\item Reactions occur over the surface of catalytic units (pelets, particles, etc.), which may be either porous or non-porous.
\item The distribution of catalytic units within the reactor space is non-uniform.
\end{itemize}

There is presently no general theory that addresses all these elements together, especially the last one, regarding the geometric configuration of the catalytic units distributed within the reactor space and separated by the non-active material. For a few references on this general topic we mention  the classic works \cite{Aris,FK} as well as \cite{YBGE}.
In the present work we propose to address this need by modeling the geometric configuration in terms of a network structure, in the context of systems of linear reactions.  The following subsections of this introduction explain our general set-up, with detailed definitions given in subsequent sections. 

 In open reactors, one specific problem of basic interest is the determination of the composition of the reactive mixture after the reactor is fully evacuated,  for a given  mixture introduced initially. 
 Such ``injection-evacuation'' operation is one of the basic procedures in chemical engineering, and it may take many forms. For example,
   a similar problem was studied within the Temporal Analysis of Products (TAP)  approach \cite{CYMGI,CYMGII,CYMGIII,CSYMG}. Also the problem analysed in \cite{GYOC},  concerning  complex catalytic reactions accompanied by deactivation, is of this kind, where catalytic deactivation may be regarded  as equivalent to reactor outflow in the present context. In such situations, a basic problem is to determine the  composition of this output mixture.  
   
In our network setting, this {\em output composition problem} (OCP)    can be analysed in 
a computationally effective way. This analysis is the main concern of the present paper.

The common theoretical approach for obtaining output composition is to begin by solving the reaction-transport equations, from which one obtains the exit flow of substances from the reactor and, through integration in time, the amounts of the reaction products. We show, however, that the output composition can be expressed directly as the solution to a boundary value problem for a time-independent system of partial differential equations, thus by-passing the technically more difficult analysis based on first solving the reaction-transport equations. In the setting of network reactors (to be introduced shortly) the boundary-value problem for the OCP reduces to a system of linear algebraic equations from which we can in many cases obtain explicit solutions in a straightforward manner. 

It has long been noted that time-integral characteristics (or moments) of reactor outflow provide important information about the reaction-transport system, such as the determination of conversion from kinetic data. Early works in this direction are by Danckwerts and Zwietering \cite{D53,D58,Z}, where the authors used non-reactive tracers for this purpose. In Danckwerts's  approach (\cite{D53,D58}), a moment-based analysis  of reactor output was used. By injecting a radioactive tracer at the reactor inlet and measuring its concentration at the outlet, Zwietering (\cite{Z}) showed that mixing patterns can be determined.  In a similar vein is  {\em Temporal Analysis of Products} (TAP) approach \cite{GYPS,GYOC,MYC,WKYF,YSCG}. In   TAP studies, an insignificant amount of chemical reactants is injected at the reactor
inlet and the resulting response at the outlet is subsequently analysed. Here again, a moment-based technique for  the calculation of time-integral characteristics of reactor outflow is used. In many TAP systems, Knudsen diffusion is the only transport mechanism.

The novelty in our approach, as already noted, is that we seek an effective method for computing integral characteristics in reaction-transport systems directly, without the need to first solve the (linear) reaction-transport equations. 

At the core of our analysis is a matrix $f(\mathbf{x})=(f_{ij}(\mathbf{x}))$, which we call the {\em output composition matrix}, where the indices $i,j$ label the substances involved in the reaction-transport process. This matrix is defined as the molar fraction of substance labeled by $j$ in the reactor output composition given that a unit amount of $i$ is initially supplied to the system at position $\mathbf{x}$ in the reactor.  
On networks, substances are injected at nodes, often denote by $n$ below. In that setting, we normally write $f(n)$. Most of the present paper is about the characterization and computation of the output composition matrix, and the mathematical justification of our method for computing it.

If the input mixture at node $n$  has composition given by the vector of fractions $\alpha=(\alpha_1, \dots, \alpha_N)$, where $N$ is the number of substances and the sum of the $\alpha_i$ is $1$, then the output composition is
$\beta=(\beta_1, \dots, \beta_N)$ such that
$$\beta_i =\sum_{j=1}^N \alpha_j f_{ji}(n). $$
Thus the output composition matrix  summarizes  all the information about the reaction-transport system that bears on the determination of the composition of output mixture, disregarding    time-dependent characteristics of the   output flow. A preliminary formulation of our main result is given  in the next subsection.

Perhaps an analogy is useful in explaining the utility of the output composition matrix $f(n)$. It plays for linear reaction-transport systems a similar role to that of scattering operators in quantum theory: we may think of the different chemical species as different scattering channels; if we probe the system by injecting $i$ at node $n$, the output in channel $j$ is given by the matrix coefficient $f_{ij}(n)$. 

The present work continues the line of investigation initiated in \cite{FWY} and \cite{FWSY}, where we considered the irreversible reaction $A\rightarrow B$ on networks and studied  reaction yield   as a function of the network configuration and reaction/diffusion coefficients. That study was   based on   the Feynman-Kac formula and stochastic analysis.  In addition to greatly extending our previous work by allowing much more general systems of reactions, we have   replaced the stochastic analysis with a more concrete and elementary, and perhaps conceptually more transparent,  approach entirely based on an analysis of the initial/boundary-value problem for the reaction-transport equations. 

Our approach admits natural generalizations that we did not want to pursue here. For example, it is possible to refine the output composition matrix so that it gives amounts of reaction products that evacuated  at different parts of the reactor exit boundary.

A computer program for the computation of output composition based on the main theoretical result of this paper has been written by the second author and is available at 
\url{https://github.com/Pasewark/Reaction-Diffusion-output-composition/tree/main}.

\subsection{General set-up and formulation of the output composition  problem}

The purpose of this work is to provide an effective answer to the following problem. Let us consider a system of   reaction-advection-diffusion equations with linear reaction rate functions, taking place inside a reactor represented by $\mathcal{R}$, which may be imagined for the moment (precise definitions will be given shortly) as a region in $3$-space partially enclosed by walls that are impermeable to the flux of gases\----to be referred to as the {\em reflecting boundary} of $\mathcal{R}$\----but still open at certain places to an exterior that is  kept at vacuum conditions.

Thus the two-dimensional boundary of $\mathcal{R}$ is the union of two parts: the reflecting
boundary and the {\em exit boundary}.
  The interior of $\mathcal{R}$ is permeable to gas transport and contains a distribution of solid catalysts  promoting first order reactions among $N$ gaseous chemical species. We designate  by $1, 2, \cdots, N$ the 
various  species and by $\kappa_{ij}(\mathbf{x})$   the kinetic reaction coefficient for the linear reaction $i\rightarrow j$ at position $\mathbf{x}$. (Notice that our reactions are, formally, a system of isomerizations, although more general types can be accommodated by our analysis as will be indicated in Subsection \ref{linear rates}.) 

We have in mind situations in which the $\kappa_{ij}(\mathbf{x})$ are nonzero only at relatively small {\em active regions}.  Diffusion coefficients and advection velocity fields are also specified for each gas species.  Given this system configuration, we  suppose  that a mixture of these gas reactants is injected into $\mathcal{R}$ at known places  and reaction products are collected at the exit boundary. The precise mode of injection does not need to be specified, although it is assumed that substance amounts are sufficiently small that the linear character of transport is   attained early on in the process. After sufficient time, the reactor empties out  and the full amount of reaction  products is collected. 

\begin{wrapfigure}{r}{0.4\textwidth}
\begin{center}
 \includegraphics[width=0.4\textwidth]{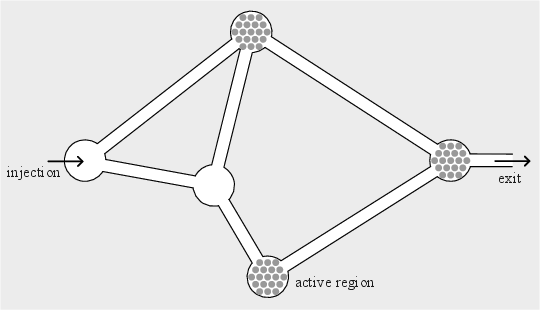}
\end{center}
\caption{\small   A network-like reactor   consists of pipes connected at junctures, which are places at which reactions may occur. The reactor interior is permeable to gas transport. Active regions contain  catalyst particles  that promote reactions involving the   gas species.}
\label{network-like reactor}
\end{wrapfigure} 

It will be shown that this {\em output composition problem}\----abbreviated OCP\----can be solved by a boundary-value problem for a system of time-independent partial differential  equations. 

We are particularly interested in reactors that have a network configuration, as suggested by Figure \ref{network-like reactor}.  That is, they consist of pipes  linked to each other at {\em junctures} of relatively small volume. The whole ensemble will be called a {\em network-like reactor}. Reactions can only take place at  junctures. Pipes   are characterized by their cross-sectional areas,  diffusivity and advection velocity, possibly dependent on species index $i$ but constant along  pipe cross-section; and the junctures are characterized by the functions
$\kappa_{ij}(\mathbf{x})$, possibly equal to  $0$.

\begin{wrapfigure}{l}{0.4\textwidth}
\begin{center}
 \includegraphics[width=0.4\textwidth]{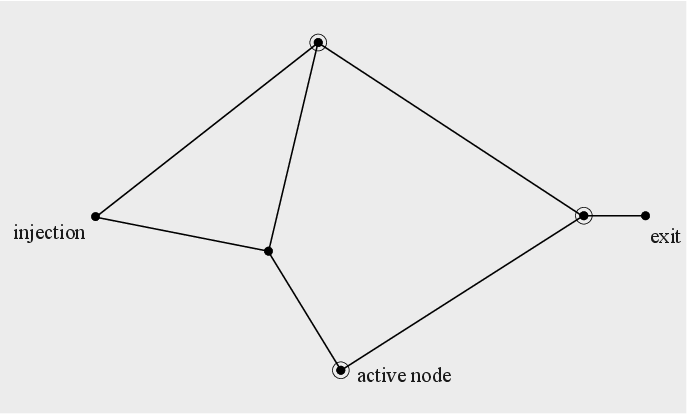}
\end{center}
\caption{\small  Network reduction of the network-like reactor of Figure \ref{network reactor}. }
\label{network reactor}
\end{wrapfigure}

Given this information, and assuming that gas concentrations in the  pipes are essentially constant along cross-sections, which is to be expected if pipes are long and narrow and diffusivity and advection velocities are also constant on cross-sections, the network-like reactor $\mathcal{R}$ becomes, effectively, $1$-dimensional (see Figure \ref{network reactor}).  In this setting, we refer to  pipes as {\em branches},  junctures as {\em nodes}, and $\mathcal{R}$ as a {\em  network reactor} (dropping the suffix ``like'').  In this network setting, we show that the boundary-value problem referred to above can be solved with relative ease.

By an effective answer to this output composition problem we have in mind formulas for the amount of each gas species in the reactor output showing  the explicit dependence on  reaction coefficients at the various active nodes, with coefficients that are known functions of the geometric/topological parameters of the network such as lengths of branches and degrees of nodes, initial data, and the coefficients of diffusivity and  advection velocities. (The latter transport parameters will be assumed constant along pipes in our examples.) 
As will be seen, the output composition problem for network reactors can be conveniently expressed in terms of a matrix $f(n)=(f_{ij}(n))$ whose entry $f_{ij}(n)$ is the fraction of species $j$ in the output given that a unit pulse of $i$ is injected into the reactor at node $n$. 
When the transport parameters are constant along branches,  this quantity will turn out to be  a rational function of the reaction rate constants with coefficients that  are polynomial functions of the {\em velocity-adjusted lengths} of the network branches, a measure of the effective length of  pipes to be defined shortly.

\subsection{Organization of the paper}
The paper is organized as follows. In Section \ref{main problem},  we describe the system of partial differential equations and initial/boundary-value conditions that serve as the mathematical model for the reaction-transport process and   show the boundary-value problem that  solves the OCP for general reactor domains. (Subsection \ref{Sub_bvp}.) We then introduce definitions and notation needed to describe the network version of the OCP (Subsection \ref{def_notat}).  In Subsection \ref{linear rates} we make a few remarks about the nature of the reactions this study is restricted to and in Subsection \ref{kappaK} we describe the relationship between reaction coefficients on the network-like reactor and on its network approximation. The matrix $f(n)$ encoding the solution to the OCP is explained in greater detail in Subsection \ref{Matrix f} and the method for obtaining it as solution of a system of linear algebraic equations is shown in Subsection \ref{solution}. It turns out that the presence of advection velocities enters into $f(n)$ in a very simple way through the introduction of {\em velocity-adjusted lengths} defined and explained in Subsection \ref{vel_adjusted}. Section \ref{examples} provides several examples to illustrate the procedure for solving the OCP and makes a few observations about the nature of solutions. In Section \ref{proof} we sketch the mathematical proof of the main result for general reactor domains and in Section \ref{reduction} we show how those results are formulated for network reactors. We end with the main conclusions in Section \ref{conclusions} and in Section \ref{symbols} we provide a glossary  of the  most frequently used symbols.

\section{Definitions and method of solution} \label{main problem}

 \subsection{The boundary-value problem solving the OCP}\label{Sub_bvp} For the moment, let us assume that the reactor $\mathcal{R}$  is a general domain in coordinate $3$-space with nice (differentiable) reflecting and exit boundaries.
The mathematical model for the reaction-transport system is as follows. Let $D_i(\mathbf{x})$ and $\mathbf{v}_i(\mathbf{x})$ represent the diffusivity and advection velocities, which   we allow to depend on  the  gas species. These species are labeled by $i=1,\dots, N$.
The concentrations $c_i(\mathbf{x},t)$   satisfy the system of equations (see, for example, \cite{Aris}) 
\begin{equation}\label{reaction_transport} \frac{\partial c_i}{\partial t}+\nabla\cdot\mathbf{j}_i=\sum_jc_j\kappa_{ji},\end{equation}
where $\mathbf{j}_i=c_i\mathbf{v}_i -D_i\nabla c_i$ is the flux vector field of species $i$,  and  boundary   conditions:
\begin{itemize}
\item $c_i(\mathbf{x},t)=0$ for $\mathbf{x}$ on the exit boundary of $\mathcal{R}$;
\item the normal component of the flux, $\mathbf{n}(\mathbf{x})\cdot \mathbf{j}_i(\mathbf{x},t)$, equals zero on the reflecting boundary.
\end{itemize}  
Here $\mathbf{n}(\mathbf{x})$ is the unit normal vector pointing outward at a boundary point $\mathbf{x}$ of $\mathcal{R}$.
One further specifies initial concentrations at time $t=0$. 

In the long run, all reaction products (including gas that didn't undergo reaction) leave   $\mathcal{R}$ through the exit boundary. The amount of each species in the reactor output is the    quantity of primary interest in our analysis. Since the initial-boundary-value problem is linear, in order to predict  the output composition it is sufficient to determine the quantities $f_{ij}(\mathbf{x})$
defined as:
\begin{equation}\label{def_fij}f_{ij}(\mathbf{x})=\text{ fraction of $j$ in the output given that a unit pulse of $i$ is initially injected at $\mathbf{x}$.} \end{equation}
We call $f(\mathbf{x}):=(f_{ij}(\mathbf{x}))$ the {\em output composition matrix} for each injection point $\mathbf{x}$. Our analysis begins with the observation  that this matrix-valued function on $\mathcal{R}$   satisfies the  system of partial differential equations
\begin{equation}\label{pde}
\nabla\cdot\left(D_i\nabla f_{ij}\right) +\mathbf{v}_i\cdot \nabla f_{ij} +\sum_k\kappa_{ik}f_{kj}=0, \ \ \ j=1, \dots, N,
\end{equation}
and boundary conditions
\begin{equation}\label{pde_bc}
\begin{aligned}
\mathbf{n}\cdot \nabla f_{ij}&=0\ \ \ \text{on the reflecting boundary of $\mathcal{R}$}\\[6pt]
f_{ij}&=\delta_{ij}\ \ \ \text{on the exit boundary of $\mathcal{R}$.}
\end{aligned}
\end{equation}
This result will be proved in Section \ref{proof}. Under a few simplifying assumptions  that are natural to  network-like reactors, it is possible to solve the boundary-value problem explicitly in many cases.  In the network approximation, the reactor domain $\mathcal{R}$ becomes a union of (possibly curved) line segments that we call {\em branches}, which are connected to each other at points called {\em nodes}.  We further assume that the transport quantities (diffusivities and advection velocities) are constant along  branches   and  chemically active regions are reduced to nodes.  The reduction of the OCP and its solution to network reactors will be detailed in Section \ref{reduction}.

We begin this analysis in the next subsection by introducing some notation and terminology for reaction-transport processes on networks.

\subsection{Some network definitions and notation}\label{def_notat} 
\begin{wrapfigure}{l}{0.6\textwidth}
\begin{center}
 \includegraphics[width=0.6\textwidth]{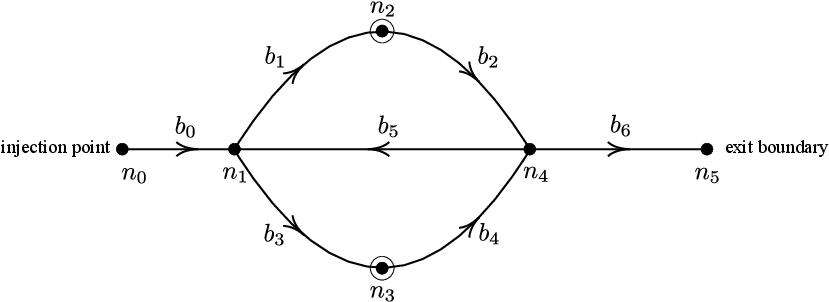}
\caption{\small   The main elements of a network reactor. Active nodes are indicated by a circle around a solid dot.}
\label{ExampleNetwork}
\end{center}
\end{wrapfigure}

As already indicated, the network-like reactor, consisting of thin pipes connected to each other at junctures, with chemically active regions restricted to junctures, will be replaced with an actual network (or graph) consisting of $1$-dimensional branches (edges) and point nodes (vertices). All the geometric, transport and reaction parameters relevant to the output composition problem will become parameters assigned to these branches and nodes. The following description refers to this network reduction, which we will call the {\em  network reactor}, or simply the {\em network}, and continue to denote by $\mathcal{R}$. Figure \ref{ExampleNetwork} will be used to  illustrate the main definitions introduced in this subsection.

Proceeding  more formally, we define a network reactor $\mathcal{R}$ as a union of finitely many lines in coordinate $3$-space with  finite length, called the reactor's {\em  branches} and indicated by the labels $b_0, b_1, \dots$. These branches are joined at points which we call the reactor's {\em nodes}, indicated by
 $n_0, n_1, \dots$. (Indexing branches and nodes starting from $0$ is, of course, an arbitrary matter. Occasionally, we begin from $1$.) Each branch $b$ can be oriented in two possible ways, indicated by $\mathbf{b}$ and $\overline{\mathbf{b}}$. It is useful to make from the beginning an arbitrary choice of orientation for each $b$ (indicated by arrows in the network diagrams) so that branch velocities can be expressed by a (positive or negative) number attached to $b$.  We may occasionally use the opposite  branch orientation than the one indicated by the arrow. This may  happen, for example, when writing node conditions for our differential equations, where   
 it is convenient to orient branches attached to a given node in the direction pointing away from the node.  In such situations, the sign of the advection velocity is flipped. An oriented branch $\mathbf{b}$ may also be indicated by the pair of its initial and terminal nodes, $(n,n')$. This can only be done when there is at most one branch between any pair of nodes. By adding additional inert nodes, this assumption can be made without any loss of generality and without changing the properties of the system. We then have $\overline{(n,n')}=(n',n)$.

To summarize, branches and nodes of the network reactor $\mathcal{R}$ are assigned the following set of parameters:
\begin{itemize}
\item 
To each branch $b$ is associated its length $\ell(b)$, diffusivity coefficients $D_i(b)$ (also indicated by $D_i^b$ at some places in the analysis) where $i=1, \cdots, N$ labels the gas species, and advection velocities $\nu_i(\mathbf{b})$. Velocities can be positive, negative or zero and we have 
$\nu_i(\overline{\mathbf{b}})=-\nu_i(\mathbf{b})$.  When  branch orientations are explicitly  indicated on network diagrams by arrows (as we do in the examples), we may simply write $\nu_i(b)$ or $\nu_i^b$. 
\item To each node $n$ we associate a matrix $K(n)=(K_{ij}(n))$ where $K_{ij}(n)$ is the constant of the $i\rightarrow j$ reaction at node $n$. It is mathematically convenient to define $K_{ii}(n)$ to be equal to the negative of the sum of the $K_{ij}(n)$ for all $j$ not equal to $i$.
Defined this way,  the sum of the entries of each row of $K(n)$ is $0$.  Notice that we are using  different symbols for the reaction coefficient $\kappa_{ij}(\mathbf{x})$ for a general (network-like) reactor and $K_{ij}(n)$ for the corresponding constant on the network reactor approximation. The latter is obtained from the former by a scaling limit and so they  are, as we will see shortly (Subsection \ref{kappaK}), distinct quantities. It is the $K_{ij}(n)$ (and not the $\kappa_{ij}(\mathbf{x})$) that will appear in our explicit formulas for output composition on network reactors. 
\item  To each node $n$ and each branch $b$ attached to $n$ we define $p(n,b)$ as the ratio of the cross-sectional area of the pipe (of the original network-like domain, represented by $b$ in the network reduction) over the sum of the cross-sectional areas of all the branches attached to $n$. So defined, the sum of the $p(n,b)$ for a given $n$ is $1$. In the examples, we always
take $p(n,b)=1/\text{deg}(n)$, where the degree of a node, $\text{deg}(n)$, is the number of branches connected to it. This amounts to the assumption that all pipes have the same cross-sectional area.
\item We may identify an oriented branch $\mathbf{b}$ with the closed interval $[0,\ell(b)]$. More precisely, we associate to $\mathbf{b}=(n,n')$ a (twice continuously differentiable) parametrization of $b$ by arc-length, $\varphi_{\mathbf{b}}(x)$, where $0\leq x\leq \ell(b)$, so that $\varphi_\mathbf{b}(0)=n$ and
$\varphi_{\mathbf{b}}(\ell(b))=n'$.  We are thus indicating points along $b$ by their distance (length) from the initial node.  If $f$ is any function on $\mathcal{R}$,
its restriction to a branch $b$ becomes a function of $x$: $f^{\mathbf{b}}(x):=f(\varphi_{\mathbf{b}}(x))$. The reason for writing $\mathbf{b}$ as a superscript 
is that such functions will typically be further indexed by the label of the chemical species.  For example, the concentration of $i$ at the point $\varphi_{\mathbf{b}}(x)$ may be written $c^{\mathbf{b}}_i(x):=c_i(\varphi_\mathbf{b}(x))$.   Notice that the derivative
$\varphi'_{\mathbf{b}}(x)$ is a unit length vector. If $\mathbf{v}_i$ is the advection velocity of $i$, then  
$\mathbf{v}_i(\varphi_{\mathbf{b}}(x))=\nu_i^{\mathbf{b}}(x)\varphi'_{\mathbf{b}}(x).$ (We are not yet assuming that advection velocities and diffusivities are constant along branches.)
\end{itemize}

A node will be called (chemically) {\em active} if  some reaction coefficient at it is positive. An  {\em inactive} node is one at which all reaction coefficients equal $0$. A subset of  inactive nodes will constitute the 
  {\em exit boundary}. In the example of Figure \ref{ExampleNetwork}, this is the single-point set consisting of $n_5$. The exit boundary is where  products can leave the reactor and where concentrations are set equal to zero (vacuum conditions).
  We call {\em internal} nodes those nodes that do not lie in the exit boundary.
  A collection of internal nodes, active or inactive,  may be chosen for the place at which an initial pulse of reactants is injected into the reactor.

  In practice, it is useful to allow the exit boundary to consist of multiple nodes but, mathematically, no generality is lost if we assume that all these exit nodes  merge into a single one. 
  It is also convenient, as already noted, to suppose that, between each pair of nodes, there can be at most one connecting branch, so that an oriented branch
  $\mathbf{b}$ can also be indicated by the pair  $(n,n')$ of initial and terminal nodes. When describing vectors (advection velocities and gradients of functions) both notations, say $\nu_i(\mathbf{b})$ or $\nu_i(n,n')$, may be used depending on convenience and context.

  \subsection{General reactions with linear rate functions} \label{linear rates}
We have assumed that reactions are linear, of the type $i\rightarrow j$,
but our analysis can accommodate more general types so long as rate functions are linear. This is explained below.

For the purposes of this subsection, we regard chemically active junctures in the network-like reactor as CSTRs with in and out fluxes where junctures connect to pipes.
Let us suppose that the reaction mechanism  on a given juncture   consists of a set of $S$ reactions\----generally an even number since we count separately a reaction and its reverse\----involving $N$ molecular species.  We denote these species  here by $A_1,\dots, A_N$. (Elsewhere in the paper, they are indicated simply by $i$ rather than $A_i$.) Let each  reaction, labeled by the index $s=1,\dots, S$ and  given by
$$ a_{s1}A_1+\cdots+a_{sN}A_N\rightarrow b_{s1}A_{1}+\cdots+b_{sN} A_N,$$
have reaction rate functions $w_s(c)$ where $c=(c_1,\dots,c_N)$. The stoichiometric coefficients, $a_{si}, b_{si}$,  are non-negative integers and $a_{si}b_{si}=0$ so that no  species    appear on both sides  of the reaction equation. The rate of change of the concentration of $A_i$ is
$$\frac{dc_j}{dt} =\sum_{s=1}^S (b_{sj}-a_{sj})w_s(c)+\text{net flux of $A_j$ in and out of juncture}.$$

We assume that the rate function $w_s(c)$ is a  linear function of   the concentration of  one of the input species (that is to say, one of the $A_i$ for which $a_{si}$ is not zero). This  approximation  is often made in two cases: (a)  the actual reaction mechanism involves a linear step whose rate coefficient is significantly smaller than those of the other reaction steps; in such a case, the overall  rate can be dominated by that of the slower linear reaction.  (b) The reaction is bimolecular  and involves components $A_1$ and $A_2$,   $A_2$ being much more   abundant than  $A_1$. In this case,   the reaction rate can often be approximated by a linear function of the concentration of $A_1$. We refer to \cite{YBGE} for more details on such issues in chemical kinetics. Generally, such linearization is an important problem in chemical engineering, but it is an issue that lies outside the scope of the present paper.

 Thus we assume that 
$w_s(c)=w_{i_s}c_{i_s}$ for some $i_s$. By introducing $$\kappa_{ij}:=\sum_{s=1}^S w_{is}(b_{sj}-a_{sj})\delta_{ii_s}$$
we reduce these differential equations to our preferred form:
$$\frac{dc_j}{dt} =\sum_{i=1}^N c_i\kappa_{ij}+ \text{net flux of $A_j$}.$$
 In our analysis of the OCP, we rely on general facts about solutions of systems of parabolic partial differential equations that  require further assumptions on the entries of the matrix $\kappa$, the most important being that the off-diagonal entries should be non-negative. (See, for example, \cite{Volp} 3.4.1.) 
Reaction equations for which these assumptions  most directly apply  are of the type $A_i\rightleftharpoons A_j$. However, it is possible to  accommodate more general types whose rates are still linear. 
Consider, for example, the situation in which species $A_1, A_2, A_3$ are involved in the  irreversible reaction
$$2A_3\rightarrow A_1+A_2$$
 with rate $w(c)=w c_3$.  In this case, $i_s=3$. We then have $\kappa_{ij}=0$ if $i\neq 3$ and the third row of $\kappa$ is $w(1,1,-2)$.  
Observe that this system is equivalent to two irreversible isomeric reactions: $A_3\rightarrow A_1$, $A_3\rightarrow A_2$. 
For another example, let us consider the system of three reaction pairs:  
$$
\ce{2A_1
<=>[w^{(1)}_+][w^{(1)}_-]{2A_3}
},\ \ \ \ 
\ce{2A_2
<=>[w^{(2)}_+][w^{(2)}_-]{2A_3}
},\ \ \ \ 
\ce{A_1
<=>[w^{(3)}_+][w^{(3)}_-]{A_2}
}.
$$
Here the rate functions are  $w_+^{(1)}c_1$ and $w_-^{(1)}c_3$ for the first pair,
$w_+^{(2)}c_2$ and $w_-^{(2)}c_3$ for the second, and $w_+^{(3)}c_1$ and $w_-^{(3)}c_2$ for the third.
In this case,
$$\kappa =  \left(\begin{array}{ccc}-\left(2w_+^{(1)}+w^{(3)}_+\right) & w^{(3)}_+ & 2w_+^{(1)} \\[6pt]
w_-^{(3)} & -\left(w_-^{(3)}+2w^{(2)}_+\right) & 2 w_+^{(2)} \\[6pt]
2w_-^{(1)} & 2w_-^{(2)} & -2\left(w_-^{(1)}+w_-^{(2)}\right)\end{array}\right). $$
Notice that the off-diagonal elements of this matrix are non-negative and the sum of the elements of each row is $0$. These are the two properties of $\kappa$ that we assume for the main results of this paper.  

It is not clear to us how general the matrix $\kappa$ can be  for our analysis to still be valid. This analysis can be significantly extended to treat  more general (multimolecular) reactions than considered so far, even if many of them may not be meaningful from a  chemical engineering perspective (although they could be relevant in other areas of science where kinetic models are used).
It is an interesting  problem to determine how general a system of  reactions with linear rates we can accept,  but this article is not the place to deal with 
 such mathematical issues. In all the examples discussed below, we have limited ourselves to  reactions of type $A_i\rightleftharpoons A_j$.

\subsection{The relationship between $\kappa_{ij}(\mathbf{x})$ and $K_{ij}(n)$}\label{kappaK}
Next, we need to understand the relationship between the reaction coefficients $\kappa_{ij}(\mathbf{x})$ on the network-like reactor in dimension $3$ and the corresponding constants $K_{ij}(n)$ defined at a node $n$ on
the network reduction. Recall that on the network-like reactors, active regions are localized but still $3$-dimensional. In the passage to network reactors these  regions shrink down to points. This entails a change of physical units,  so that $K_{ij}(n)$ has units of distance over time,  not the reciprocal of time. The purpose of the present section is to provide some clarification on this issue.

When we get to see explicit formulas  for output composition later in the paper, we will often encounter dimensionless expressions of the form
$\ell K/D$, where $\ell$ is a length, $D$ a constant of diffusivity, and $K$ a reaction coefficient associated with an active node. This may seem at first to conflict with standard quantities in chemical engineering. 
Notice that $\kappa$, with the physical unit   reciprocal of time, is a sort of ``density of chemical activity.'' If this activity is strongly localized along a small interval of length $\epsilon$ on a network branch, one needs to integrate over this length to obtain  activity on that small interval. Then $\kappa$ should scale as $\kappa_\epsilon \approx K/\epsilon$ and
$\ell K/D \approx \ell \epsilon \kappa_\epsilon/D$.

\begin{figure}[htbp]
\begin{center}
 \includegraphics[width=4.0in]{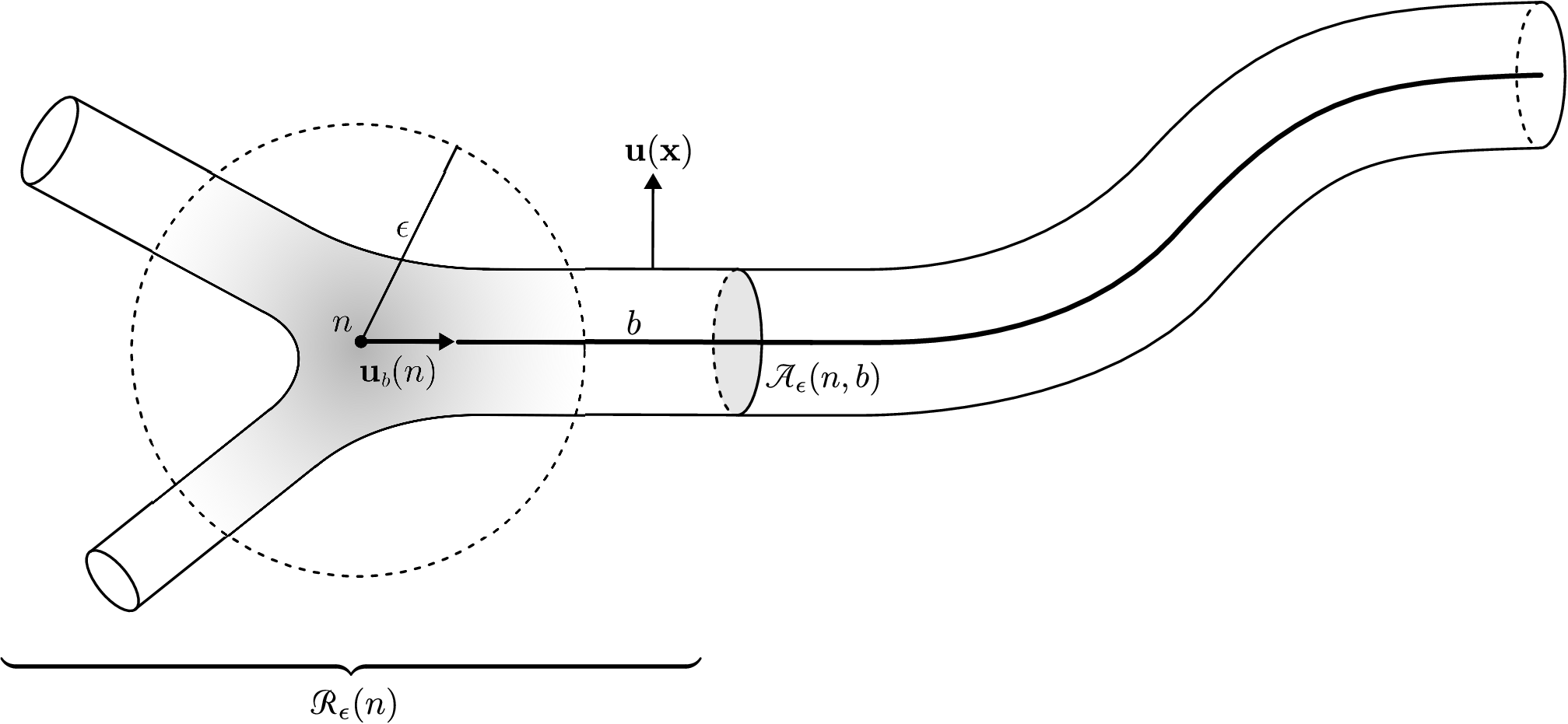}
\caption{\small  A juncture $\mathcal{R}_\epsilon(n)$ in the network-like reactor corresponding to a node $n$ with an attached pipe. The gradient shading   represents catalyst distribution and  $\epsilon$ is a radius delimiting the chemically active region. The unit vector $\mathbf{u}(\mathbf{x})$ is orthogonal to the {\em reflecting boundary} of $\mathcal{R}$ whereas $\mathbf{u}_b(n)$ is the unit vector orthogonal to the cross-section of the pipe   $b$ pointing away from $n$. 
$\mathcal{A}_{\epsilon}(n,b)$ is the cross-sectional disc where the pipe   attaches to the juncture. Other features are explained in the text.}
\label{Juncture}
\end{center}
\end{figure}

More precisely, let us look at what happens in a neighborhood of an active juncture.
The network  approximation of the network-like reactor made of pipes and junctures will depend on a small length parameter $\epsilon$. This is   the radius of a ball centered at any node $n$ that delimits the  chemically active region at that juncture, if the juncture is active, and the part of the network near $n$ having a relatively complicated geometry, as opposed to the  simple tube-like shapes on the complement of the union of junctures. 
 See    Figure \ref{Juncture}. 
 Outside of such  balls, pipes have uniform cross-section and no reactions take place. This same $\epsilon$ is assumed to hold for all the nodes  and the  smaller the value of $\epsilon$ the better is the network approximation.

 Since, in the network model, active regions   shrink to points, the reaction coefficients  $\kappa_{ij}(\mathbf{x})$ at any position $\mathbf{x}$ should be scaled up with the reciprocal of $\epsilon$.  This is because the probability that a molecule will react is proportional to the total amount of time it spends in the active regions; as these regions shrink to a point, the rate constants should scale up accordingly.
 We indicate this by writing   $\kappa_{ij}(\epsilon, \mathbf{x})$.

Let $V_\epsilon(n)$ and $A_\epsilon(n)$ be, respectively, the volume of the juncture $\mathcal{R}_\epsilon(n)$ and the area of the part of the  boundary of this juncture  complementary to the reflecting boundary\----a union of pipe cross-sectional discs. In taking the limit as $\epsilon$ approaches $0$, we suppose that the dimensionless quantity
$ \epsilon A_\epsilon(n)/V_\epsilon(n)$ converges to the positive quantity $\chi(n)$, a geometric characteristic of the juncture that survives the limit process, and  
$$\frac{1}{A_\epsilon(n)} \int_{\mathcal{R}_\epsilon(n)} \kappa_{ij}(\epsilon,\mathbf{x})\, dV(\mathbf{x})\rightarrow K_{ij}(n). $$
In the above volume integral, keep in mind that $\kappa_{ij}(\epsilon, \mathbf{x})$ is of the order $1/\epsilon$ and  has physical dimension $1/\text{time}$. Thus $K_{ij}(n)$ has physical dimension $\text{length}/\text{time}$.

Also note the limit
$A_\epsilon(n,b)/A_\epsilon(n)\rightarrow p(n,b)$, where $A_\epsilon(n,b)$\----the area of $\mathcal{A}_{\epsilon}(n,b)$ (see Figure \ref{Juncture})\----is the cross-sectional area of the pipe corresponding to the branch $b$ attached to $n$.

These considerations will become important in Section \ref{node conditions}, where we obtain node conditions for the composition output boundary-value problem.

\subsection{The boundary-value problem for output composition}
Our main result for network reactors, described in the next subsection, is a consequence of the already mentioned (see section \ref{Sub_bvp})  observation that  the output composition matrix $f(\mathbf{x})$, for a not necessarily network-like reactor domain $\mathcal{R}$, is the solution to the following boundary-value problem:  
\begin{equation}\label{pde_bis}
(\mathcal{A} f)_{ij}:=\nabla\cdot\left(D_i\nabla f_{ij}\right) +\mathbf{v}_i\cdot \nabla f_{ij} +\sum_k\kappa_{ik}f_{kj}=0, \ \ \ i,j=1, \dots, N,
\end{equation}
and boundary conditions
\begin{equation}\label{pde_bc_bis}
\begin{aligned}
\mathbf{n}\cdot \nabla f_{ij}&=0\ \ \ \text{on the reflecting boundary of $\mathcal{R}$}\\[6pt]
f_{ij}&=\delta_{ij}\ \ \ \text{on the exit boundary of $\mathcal{R}$.}
\end{aligned}
\end{equation}

To begin to develop an understanding of this boundary-value problem, let us consider two extreme cases: (a) the reaction coefficients are negligible compared  to the transport coefficients; (b) the network-like reactor consists of one small active region   directly open to the outside. In this case, there are only to positions to consider: inside and outside the reactor and  the only relevant transport characteristic is the rate of evaculation of each substance. 

In case (a), neglecting  reactions ($\kappa_{ij}(\mathbf{x})=0$),  the solution to the boundary-value  problem is the constant matrix with elements $f_{ij}(\mathbf{x})=\delta_{ij}$. The obvious interpretation is that the composition of the output mixture equals the composition of the initially injected mixture.

\begin{figure}[htbp]
\begin{center}
\includegraphics[width=1.5in]{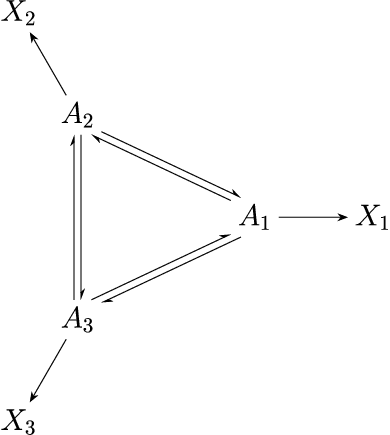}
\caption{\small A system of reactions with three terminal species (absorbing states). }
\label{just reaction}
\end{center}
\end{figure}

Let us now turn to special case (b). We don't have, in this case, the transport terms of  the general reaction-transport equation, so it is necessary to introduce a rate of evacuation. We model this situation by imagining that space consists of two points, one representing the active region and the other the outside of the reactor (the surrounding of the active region). We then introduce irreversible reactions of type $A_i\mapsto X_i$, with a rate $\mu_i$, where $A_i$ represents a substance inside the reactor and $X_i$ is the same substance outside, where $i=1, \dots, N$. Thus
this ``reaction-evacuation'' process is mathematically  equivalent to a reactions-only process inside a closed reactor (which we may imagine as a batch reactor) observed from time $0$ until chemical equilibrium is reached and only the substances $X_i$ remain. 
The matrix $f$ for the reactions-only process has the form
$$ f=\left(\begin{array}{cc}f_{AA} & f_{AX} \\f_{XA} & f_{XX}\end{array}\right)=\left(\begin{array}{cc}O & f_{AX} \\ O & I\end{array}\right),$$
where the blocks have size $N\times N$, $O$ is the zero matrix, and $I$ is the identity. The notation for the   block indices  should be interpreted as follows: the $(i,j)$-element of $f_{AX}$ gives the fraction of   $X_j$ at equilibrium when a unit amount of $A_i$ is introduced initially. With similar  interpretations, it is clear that $f_{AA}=f_{XA}=O$ and $f_{XX}=I$.

The reaction coefficients can be similarly written in block matrix form:
$$ \kappa = \left(\begin{array}{cc}\kappa_{AA} & \kappa_{AX} \\ \kappa_{XA} & \kappa_{XX}\end{array}\right) = \left(\begin{array}{cc}\kappa_{AA} & \kappa_{AX} \\ O & O\end{array}\right),$$
where $\kappa_{AX}$ is the diagonal matrix $\text{diag}(\mu_1, \dots, \mu_N)$.
The concentrations $c_i$ evolve in time according to the system of equations 
$$\frac{d c_i}{dt} = \sum_{j} c_j \kappa_{ji}.$$

We wish to show that our boundary-value problem, in this special case, is indeed solved by the output composition matrix $f$. First observe that, in the absence of diffusion and advection, $\mathcal{A}f=\kappa f$, so   Equation (\ref{pde_bis}) reduces to $\kappa f=0$, and the boundary condition (\ref{pde_bc_bis}) reduces to $f_{XX}=I$. Written in terms of the matrix blocks, these two equations amount to
\begin{equation}\label{reduced_equation}\kappa_{AA} f_{AX} +\kappa_{AX}=0.\end{equation}

In order to see that Equation (\ref{reduced_equation}) is, indeed, satisfied by $f$, we now  introduce a key ingredient  of the general proof that, in this special case (b), reduces to 
$$\rho_{ij}(t|s)=\text{ concentration of $i$ at time $t$ given that unit amount of $j$ is introduced at time $s<t$}.$$ Naturally,
$$\rho'_{ij}(t|0) = \sum_{k}\rho_{kj}(t|0)\kappa_{ki}, \ \ \rho_{ij}(0|0)=\delta_{ij}, $$
where $\rho'_{ij}(t|0)$ is derivative in $t$.
We then have the following fundamental equation, which we accept here on heuristic grounds:
\begin{equation}\label{fundamental_1}f_{ij} = \sum_{k} \rho_{ki}(t|0) f_{kj}.\end{equation}
  In words, if a unit amount of $i$ is introduced at time $0$, then the final amount of $j$ equals the final amount of $j$  given that a unit amount of $k$ is introduced at time $t$ weighted by the concentration of $k$ at time $t$. Said yet differently, the output amount of $j$ is the same whether we start with the composition consisting only of $i$ at time $0$, or with the mixture defined by $(\rho_{1i}(t|0), \dots, \rho_{Ni}(t|0))$ at time $t$. 
  
  Differentiating in $t$ at time $0$ and using the equation and initial condition for $\rho_{ki}(t|0)$ yields
  $$0=\sum_{k}\kappa_{ki}f_{kj}.$$
In this system of equations, $i$ represents one of the $A$-species and $j$ one of the $X$-species. Breaking the sum into these two types of indices, we obtain the system of equations (\ref{reduced_equation}), which is what we wished to demonstrate.

The above remark brings into consideration the matrix-valued function $\rho$. 
  Back to  general reaction-advection-diffusion systems, a similar function plays a central role in the proof given in Section \ref{proof}. The following remarks highlight the role of this key ingredient.

  First notice that the total amount   of substance $i$ produced in the long run by the reaction-transport process in the open reactor $\mathcal{R}$ is (here $dx$ is ordinary volume element):
  $$\text{Amount of $i$ in gas output } = \text{Initial amount of $i$ }+ \underbrace{\sum_j \int_0^\infty \int_{\mathcal{R}} c_j(\mathbf{x},t) \kappa_{ji}(\mathbf{x})\, dx\, dt}_{\text{net amount of $i$ produced by reactions}}.$$
We wish to view this quantity as a function of the initial mixture pulsed into the reactor at time $0$. Let us introduce the {\em fundamental solution} to the reaction-transport equations:
$$\rho_{ij}(\mathbf{x},t|\mathbf{y},s)=\text{ concentration of $i$ at $(\mathbf{x},t)$  given that a unit pulse of $j$ is injected at $(\mathbf{y},s), s<t$.}$$ Then, using the previous balance equation, we obtain that the total amount of $j$ in the gas output given that a unit of $i$ is introduced at the beginning of the process is given by
$$f_{ij}(\mathbf{y})= \delta_{ij} +\sum_k \int_0^\infty \left[\int_{\mathcal{R}} \rho_{ki}(\mathbf{x},t|\mathbf{y},0)\kappa_{kj}(\mathbf{x}) \, dx\right]\, dt. $$
In matrix form,
$$f(\mathbf{y})=I+\int_0^\infty \int_{\mathcal{R}} \kappa^\intercal(\mathbf{x})\rho(\mathbf{x,t}|\mathbf{y},0)\, dx\, dt,$$
where $\kappa^\intercal$ indicates matrix transpose.
One is led to ask for a boundary-value problem characterizing the time-independent function $f(\mathbf{y})$. We know that, as a function of $(\mathbf{x}, t)$ with $(\mathbf{y},s)$ fixed, the quantity  $\rho(\mathbf{x,t}|\mathbf{y},0)$ satisfies the reaction-transport equation with initial pulse condition $$\rho_{ij}(\mathbf{x},t|\mathbf{y},s)\overset{t\downarrow s}{\longrightarrow}\delta(\mathbf{x}-\mathbf{y})\delta_{ij}.$$
The key fact we need, well-known in the theory of parabolic differential equations (see, for example, \cite{Friedman}), is that $\rho(\mathbf{x,t}|\mathbf{y},s)$, as a function of $(\mathbf{y},s)$ for $(\mathbf{x},t)$ fixed, satisfies a similar initial-boundary-value problem. Specifically,
writing $\rho^*(\mathbf{y},s|\mathbf{x},t)=\rho(\mathbf{x},t|\mathbf{y},s)^\intercal$,  
$$-\frac{\partial \rho^*}{\partial s}=\mathcal{A}\rho^*.$$
Here,  the derivatives involved in $\mathcal{A}$ are relative to $\mathbf{y}$. (See Section \ref{relation} for details.)  This is the place where the operator $\mathcal{A}$ finally enters the picture. The rest of the verification of the main claim of this section now follows from the relatively straightforward mathematical manipulations described in greater detail in  Section \ref{proof}. Details apart, we believe the conceptual core of this story lies in the fundamental identity
\begin{equation}\label{fundamental_2}f_{ij}(x) =\sum_{k}\int_\mathcal{R} \rho_{ki}(y,t|x,0)f_{kj}(y)\, dy,\end{equation}
generalizing Equation (\ref{fundamental_1}),   which is very natural given the interpretation of the matrix $\rho(y,t|x,0)$.

\subsection{The output composition matrix for network   systems}\label{Matrix f}

By the {\em output composition matrix} for network reactors,  $f(n)=(f_{ij}(n))$, 
  we mean
a matrix-valued function of the node $n$ having the following interpretation:
$f_{ij}(n)$ is the fraction of gas species $j$ in the reactor's output given that a unit pulse of species $i$ is initially injected at node $n$. By definition,
if $n$ is an exit node, then $f(n)=I$ is the identity matrix; that is, $f_{ij}(n)=\delta_{ij}$, the Kronecker delta, which is $0$ if $i\neq j$ and $1$ otherwise.  

Due to the linearity of the system of equations giving $f(n)$, if a mixture of gases is injected at $n$ having composition vector $\alpha=(\alpha_1, \dots, \alpha_N)$ where $\alpha_i$ is the molar  fraction of $i$, then 
$\beta = \alpha f(n)$ is the vector of molar fractions in the reactor output. The fraction of $j$ in the output composition is then
$$\beta_j = \sum_{i=1}^N \alpha_i f_{ij}(n). $$ 
More generally, if the fraction $\alpha_i$ of $i$ is injected at node $n_i$, then $\beta_j = \sum_{i=1}^N \alpha_if_{ij}(n_i)$.

The determination of $f(n)$ is our  central problem. It will be shown  in Section \ref{proof}, for general reactor domains in $3$-space, that this matrix-valued function satisfies a boundary value problem for a time-independent system of elliptic differential operators which, when reduced to network domains (in Section \ref{reduction}), amounts to Equations (\ref{branch_eq}), (\ref{int_n_cond}) and (\ref{exit_n}) given below.

Summarizing the main result, the matrix $f(n)$, for each internal node $n$ of the network reactor, is obtained as the solution to the following time-independent boundary-value problem (the network counterpart of Equations (\ref{pde}) and boundary conditions (\ref{pde_bc})). On each branch $b$:
\begin{equation}\label{branch_eq}
D_i(b)f''_{ij}(x)+\nu_i(b) f'_{ij}(x)=0.
\end{equation}
Here $f'(x)$ indicates derivative with respect to the arc-length parameter $x$ along $b$. On internal nodes,
\begin{equation}\label{int_n_cond}
\sum_{b\sim n} p(n,b)D_i(b) f'_{ij}(n,b) +\sum_k K_{ik}(n) f_{kj}(n)=0,
\end{equation}
where $b\sim n$ indicates that the sum is over those branches that are attached to node $n$, and $f'_{ij}(n,b)$ denotes the derivative at $0$ of the restriction of $f_{ij}$ to 
branch $b$ in   the arc-length parameter $x$ of $b$ oriented away from $n$ (that is, so that $x=0$ corresponds to $n$). Finally, on exit nodes $n_{\tiny exit}$,
\begin{equation}\label{exit_n}
f_{ij}(n_{\text{\tiny exit}})=\delta_{ij}.
\end{equation}
Equations (\ref{branch_eq}) together with boundary conditions (\ref{int_n_cond}) and (\ref{exit_n}) are our fundamental equations  for the OCP on network reactors. 
Being a finite dimensional  system of  algebraic equations, they are easily solved by elementary means.
 (Observe that we are at this point assuming that  $D_i(b)$ and $\nu_i(b)$ are constant on branches.) If  $\nu_i(b)=0$   but allow $D_i(x)$ to vary along $b$, it is possible to reparametrize $b$ so as to make $D_i=1$. For simplicity, we assume that $D_i$ is already constant on branches.

In the remaining of this section, we rewrite the linear system of algebraic equations satisfied by $f(n)$ in a convenient form, highlighting  a useful concept which we call {\em velocity-adjusted length}. In the next section (Section \ref{examples}), we give several examples to illustrate how $f(n)$ is obtained explicitly.

\subsection{Velocity-adjusted lengths}\label{vel_adjusted}  Equation (\ref{branch_eq}) can be readily solved by elementary means. Let  $b$ be a branch attached to $n$ having length $\ell(b)$, and $x\in [0,\ell(b)]$  the arc-length parameter along $b$. We choose the parametrization that orients $b$ away from $n$ (so that $n$ corresponds to $x=0$). Recall that $F_{ij}(n,b):=f'_{ij}(n,b)$ represents the derivative at $x=0$ of the restriction of $f_{ij}$ to $b$. To be more explicit, we write $\mathbf{b}=(n,n')$. Then
\begin{equation}\label{f(x)}f_{ij}(x)=f_{ij}(n) + \frac{1-\exp\left\{-\frac{\nu_i(n,n')}{D_i(b)}x\right\}}{\nu_i(n,n')/D_i(b)}F_{ij}(n,b).\end{equation}
In particular,
\begin{equation}\label{F}
F_{ij}(n,b)=F_{ij}(n,n')=\frac{f_{ij}(n')-f_{ij}(n)}{\tilde{\ell}_i(n,n')},
\end{equation}
where we have introduced the quantity
$$\tilde{\ell}_i(\mathbf{b})=\tilde{\ell}_i(n,n'):=\begin{cases}\frac{1-\exp\left\{-{\ell(b)\nu_i(n,n')}/{D_i(b)}\right\}}{\nu_i(n,n')/D_i(b)} & \text{ if } \nu_i(n,n')\neq 0\\[6pt]
\ell(b) & \text{ if } \nu_i(n,n')=0.
\end{cases} $$
This positive quantity has physical dimension of length and it is continuous in $\nu_i(\mathbf{b})$, which is to say that $\tilde{\ell}_i(\mathbf{b})\rightarrow \ell(b)$ when the advection velocity  approaches $0$.   We refer to 
$\tilde{\ell}_i(\mathbf{b})$ as the (oriented) {\em velocity-adjusted length} of the branch $b$. Observe that $\tilde{\ell}(\mathbf{b})$ is always positive and depends on the orientation of the branch: if  $\nu_i(\mathbf{b})$ is positive, transport of $i$ in the direction of $\mathbf{b}$ is faster, and the adjusted length of $b$ is less than $\ell(b)$ while transport of $i$ in the direction of $\overline{\mathbf{b}}$ is slower, and the adjusted length of $b$ is greater than $\ell(b)$. Strictly speaking, this adjusted length also depends on diffusivity. Large diffusivity negates the effect of velocity by making the quotient $\nu_i(\mathbf{b})/D_i(b)$ smaller in absolute value without changing its sign, while small diffusivity accentuates the velocity adjustment. 
Introducing the dimensionless quantity  $s_i(\mathbf{b})=\ell(b)\nu_i(\mathbf{b})/D_i(b)$, we can write the above relation in more transparent form:
$$\tilde{\ell}_i(\mathbf{b})/\ell(b)=\frac{1-e^{-s_i(\mathbf{b})}}{s_i(\mathbf{b})}.$$
This function is positive, equals $1$ at $s_i(\mathbf{b})=0$, decreases to $0$ at the rate $1/s_i(\mathbf{b})$ as $s_i(\mathbf{b})\rightarrow\infty$ and grows exponentially to $+\infty$ as $s_i(\mathbf{b})\rightarrow  -\infty$. 

In our solution to the OCP, lengths of branches will always appear as $\tilde{\ell}_i(\mathbf{b})$ (or $\ell(b)$, when $\nu_i(\mathbf{b})=0$). The velocity-adjusted length is, thus, an   effective length of  branches resulting from a competition between velocity and diffusivity.

\subsection{Solution to the output composition problem on network reactors}\label{solution}
Given the observations of the previous subsection,  the entries $f_{ij}(n)$ of the output composition matrix on internal nodes $n$ are now obtained directly from 
Equations (\ref{branch_eq}),  (\ref{int_n_cond}), (\ref{exit_n}) and  (\ref{f(x)}) as solution to an ordinary linear system of algebraic equations.   To make this linear system more easily readable, it helps to   introduce the following quantities.
Let $n$ be a node and $b=(n,n')$ a branch attached to $n$. Keep in mind that the notations  $(n,b)$ and $(n,n')$ both represent an oriented
branch $\mathbf{b}$ with initial node $n$. For each $i$ we define 
 $$\xi_i(n,b)=\xi_i(n,n'):=\frac{p(n,b)D_i(b)}{\tilde{\ell}_i(n,n')}, \ \
 \eta_i(n,b)=\eta_i(n,n')=\frac{\xi_i(n,n')}{\sum_{b'}\xi_i(n,b')}$$ 
 where the sum  in the denominator of $\eta_i(n,b)$ is over all branches $b'$  attached to $n$.
 Then $\xi_i(n,b)$ has physical dimension $\text{length}/\text{time}$ and $\eta_i(n,b)$ is dimensionless. Let $\eta(n,b)=\text{diag}(\eta_1(n,b),\cdots, \eta_N(n,b))$, an $N\times N$ diagonal matrix.  Finally, we introduce the dimensionless reaction coefficients
 $$\tilde{K}_{ij}(n) :=\frac{K_{ij}(n)}{\sum_{b'}\xi_i(n,b')} $$
 where the sum is over all the branches $b'$ connected to $n$.

 We are now ready to write down the linear system for $f_{ij}(n)$. Suppose that the network contains $L+1$   nodes so that  $n_1, n_2, \dots, n_L$ are  the internal nodes and $n_{L+1}=n_{\text{\tiny exit}}$ is the exit node. We define a matrix $\Lambda$ of size $NL\times NL$, which we write in block-form, with blocks of size $N\times N$, as follows. For each pair $n_i,n_j$ of distinct internal nodes, the $N\times N$ block  $\Lambda(n_i,n_j)$ of $\Lambda$ at row $i$ and column $j$ is 
\begin{equation}\label{Lambda_def}\Lambda(n_i,n_j) =\begin{cases}
 \eta(n_i,n_j) & \text{ if } i\neq j\\[6pt]
 \tilde{K}(n_i)-I & \text{ if } i=j.
 \end{cases} \end{equation}
It is implicit in the above expression that $ \tilde{K}(n_i)=0$ if the interior node $n_i$ is not active and $\eta(n_i, n_j)=0$ if $(n_i,n_j)$ (or its opposite) is not a branch of the network. Let $$f=\left(\begin{array}{c}f(n_1) \\\vdots \\f(n_L)\end{array}\right), \ \  \lambda = \left(\begin{array}{c}-\eta(n_1,n_{\text{\tiny exit}}) \\\vdots \\-\eta(n_L,n_{\text{\tiny exit}})\end{array}\right).$$ These are $NL\times N$-sized matrices written in   block-form.
Then the output composition matrix is the solution to the linear system:  
\begin{equation}\label{fundamental} 
\Lambda f = \lambda.
\end{equation}
Explicitly,
\begin{equation}\label{matrix}\left(\begin{array}{cccc}\tilde{K}(n_1)-I & \eta(n_1,n_2) & \hdots & \eta(n_1,n_L) \\[6pt]\eta(n_2,n_1) & \tilde{K}(n_2)-I & \hdots & \eta(n_2,n_L) \\[6pt]\vdots & \vdots & \ddots & \vdots \\[6pt]\eta(n_L,n_1) & \eta(n_L,n_2) & \hdots & \tilde{K}(n_L)-I\end{array}\right)\left(\begin{array}{c}f(n_1) \\[6pt]f(n_2) \\[6pt]\vdots \\[6pt] f(n_L)\end{array}\right)=- \left(\begin{array}{c}\eta(n_1,n_{\text{\tiny exit}}) \\[6pt]\eta(n_2,n_{\text{\tiny exit}}) \\[6pt]\vdots \\[6pt] \eta(n_L,n_{\text{\tiny exit}})\end{array}\right)\end{equation}
with blocks of size $N\times N$, where $N$ is the number of gas species. 
This is our fundamental system of equations.

As an example, for the network diagram of Figure \ref{ExampleNetwork} (in which the interior nodes are $n_0, \dots, n_4$), this system becomes (notice that $\eta(n_i,n_j)=I$ if there is a single branch issuing from $n_i$)
\begin{equation}\label{big}\left(\begin{array}{ccccc}-I & I & 0 & 0 & 0 \\[6pt]0 & -I & \eta(n_1,n_2) & \eta(n_1,n_3) & \eta(n_1,n_4) \\[6pt]0 & \eta(n_2,n_1) & \tilde{K}(n_2)-I & 0 & \eta(n_2,n_4) \\[6pt]0 & \eta(n_3,n_1) & 0 & \tilde{K}(n_3)-I & \eta(n_3,n_4) \\[6pt]0 & \eta(n_4,n_1) & \eta(n_4,n_2) & \eta(n_4,n_3) & -I\end{array}\right) \left(\begin{array}{c}f(n_0) \\[6pt]f(n_1) \\[6pt]f(n_2) \\[6pt]f(n_3) \\[6pt]f(n_4)\end{array}\right)=-\left(\begin{array}{c} 0 \\[6pt] 0 \\[6pt] 0 \\[6pt] 0\\[6pt]\eta(n_4,n_5)\end{array}\right).\end{equation}

Before exploring Equation (\ref{matrix}) further, it is natural to ask whether the coefficient matrix $\Lambda$ is indeed invertible. This is to be expected  since this linear system arises from a boundary-value problem for a system of partial differential equations whose solutions are uniquely determined. Nevertheless, it is reassuring to be able  to ascertain  solvability independently by elementary means under very general assumptions. This point is discussed in Subsection \ref{invertibility}.

\section{Examples and observations} \label{examples}
In all the examples to be considered in this paper we make the following convenient but natural assumptions:  $p(n,b)=1/\text{deg}(n)$ (all pipes have the same cross-section), $D_i(b)=D$ does not depend on the chemical species and the branch, and $\nu_i(\mathbf{b})=\nu(\mathbf{b})$ is the same for all species but may depend on the branch.  Thus, for a choice $\mathbf{b}=(n,n')$ of orientation for $b$, 
$$\tilde{\ell}(\mathbf{b})=\frac{1-e^{-\nu(\mathbf{b})\ell(b)/D}}{\nu(\mathbf{b})/D}, \ \ \xi(n,n')=\frac{D/\tilde{\ell}(\mathbf{b})}{\text{deg}(n)}, \ \  \eta(n,n')=\frac{\xi(n,n')}{\sum_{n''} \xi(n,n'')}$$
do not depend on $i$. The sum in the denominator of $\eta(n,n')$ is over all branches $(n,n'')$ attached to $n$.
Since $\eta(n,n')$ can now be viewed as a scalar, we write  $\eta(n,n')I$ for the corresponding $N\times N$ matrix, where $I$ is the  identity matrix. When the branches are indexed, $b_i$, it is useful to write $\tilde{\ell}_{i}:=\tilde{\ell}(\mathbf{b}_i)$ and $\tilde{\ell}_{\overline{i}}:=\tilde{\ell}(\overline{\mathbf{b}}_i)$, where $\mathbf{b}_i$ has the orientation indicated by an arrow in the network  diagram of each example.

\subsection{Segment reactor with $1$ active node}
\begin{wrapfigure}{r}{0.4\textwidth}
\begin{center}
 \includegraphics[width=0.38\textwidth]{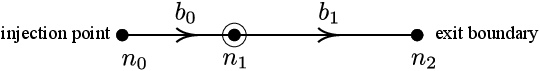}
\end{center}
\caption{\small  A simple network reactor with a single active node.  }
\label{ExampleNetwork0}
\end{wrapfigure}

For the example of Figure \ref{ExampleNetwork0}, we have 
$$\tilde{\ell}_0:=\tilde{\ell}(\overline{b}_0)=\frac{e^{\nu(b_0)\ell(b_0)/D}-1}{\nu(b_0)/D}, \ \ \tilde{\ell}_1:=\tilde{\ell}(b_1)=\frac{1-e^{-\nu(b_1)\ell(b_1)/D}}{\nu(b_1)/D},$$
so that 
$$
\eta(n_1,n_0)=\frac{\tilde{\ell}_1}{\tilde{\ell}_0+\tilde{\ell}_1}, \ \  \eta(n_1,n_2)=\frac{\tilde{\ell}_0}{\tilde{\ell}_0+\tilde{\ell}_1}, \ \ \tilde{K}:=\tilde{K}(n_1)= \frac{K(n_1)}{\frac12D\left(\frac1{\tilde{\ell}_0}+\frac1{\tilde{\ell}_1}\right)}=\frac{2\tilde{\ell}_0\tilde{\ell}_1}{\tilde{\ell}_0+\tilde{\ell}_1}\frac{K(n_1)}{D}. $$ 
Therefore,
$$\left(\begin{array}{cc}-I & I \\[6pt] \frac{\tilde{\ell}_1}{\tilde{\ell}_0+\tilde{\ell}_1}I & \tilde{K}-I\end{array}\right) \left(\begin{array}{c}f(n_0) \\[6pt]f(n_1)\end{array}\right)=-\left(\begin{array}{c}0 \\[6pt]\frac{\tilde{\ell}_0}{\tilde{\ell}_0+\tilde{\ell}_1}I\end{array}\right).$$

This system is easily solved. The result is
\begin{equation}\label{equationf}f(n_0)=f(n_1) = \left(I-\frac{2\tilde{\ell}_1}{D} K\right)^{-1}. \end{equation}
As could have been expected,   $\tilde{\ell}_0$ does not appear in the solution. Had we assumed $\ell_0=0$, however, the factor of $2$ in front of $\tilde{\ell}_1$ would be $1$ instead, since the degree of $n_1$ would go from $2$ to $1$.

Let us explore this solution in some detail. 
Recall that $f_{ij}(n_0)$ is the fraction of $j$ in the output given that a unit pulse of $i$ is initially injected at node $n_0$. 
Suppose there are only two chemical species, denoted $1, 2$ with reactions $1\rightleftharpoons 2$ so that 
$$ K=\left(\begin{array}{rr}-k_+ & k_+ \\[6pt]k_- & -k_-\end{array}\right).$$
($k_+$ is the coefficient of $1\rightarrow 2$ and $k_-$ is the coefficient for the reverse reaction $2\rightarrow 1$.) The inverse matrix in Equation (\ref{equationf}) is easily found:
$$f(n_0) =\frac{1}{1+\frac{2\tilde{\ell}_1}{D}\left(k_-+k_+\right)} \left(\begin{array}{cc}1+\frac{2\tilde{\ell}_1k_-}{D} & \frac{2\tilde{\ell}_1k_+}{D} \\[6pt]\frac{2\tilde{\ell}_1k_-}{D} & 1+\frac{2\tilde{\ell}_1k_+}{D}\end{array}\right). $$
Thus, for example, the fraction of species $2$ in the output given that a pulse containing only species $1$ was initially injected at $n_0$ is
$$f_{12}(n_0)=\frac{\frac{2\tilde{\ell}_1}{D}k_+}{1+\frac{2\tilde{\ell}_1}{D}\left(k_-+k_+\right)}. $$

From $f(n_0)$ we can determine the output composition for any   composition of substances initially injected into the reactor. Suppose that the initial quantities  of $1$ and $2$ are given by the vector $\alpha=(\alpha_1, \alpha_2)$. Then the output amounts are given by the vector $\beta=(\beta_1,\beta_2)$ such that $\beta=\alpha f(n_0)$ (matrix multiplication). In particular, 
$$\frac{\beta_2}{\beta_1}=
\left[
{k_+ +\frac{D}{2\tilde{\ell}_1}\frac{\alpha_2}{\alpha_1+\alpha_2}}\right]\bigg/\left[{k_- +\frac{D}{2\tilde{\ell}_1}\frac{\alpha_1}{\alpha_1+\alpha_2}}\right].$$

Observe the effect of varying  the transport coefficient
  $D/\tilde{\ell}_1$. If this coefficient is small, the above ratio is approximately
  $$\frac{\beta_2}{\beta_1}\approx \frac{k_+}{k_-}.$$
This is the equilibrium value for a closed reactor.
Such situation arises when the diffusion coefficient is very small or $b_1$ is very long. Equivalently, this approximation holds when the advection velocity $\nu(b_1)$ is negative with large absolute value. 
The quantity $2\tilde{\ell}_1/D$ may be viewed as a measure of the time spent at the active node.  It has physical dimension $\text{time}/\text{length}$ rather than $\text{time}$  for the same reason that $k$ does not have dimension $1/\text{time}$ when the active region reduces to a point. (The actual time spent at a single point is $0$.)

It is also interesting to observe in this example that, independently of the transport characteristics,
$$\frac{\text{output fraction of $2$ given initial unit pulse of $1$}}{\text{output fraction of $1$ given initial unit pulse of $2$}}=\frac{f_{12}(n_0)}{f_{21}(n_0)} =\frac{k_+}{k_-}=\text{ equilibrium ratio}. $$
We refer to \cite{YCM} and \cite{YBMC} for the significance of   a reciprocity relation  of similar kind but   in a different setting.  

\subsection{Distributed input}
\begin{wrapfigure}{r}{0.3\textwidth}
\begin{center}
 \includegraphics[width=0.28\textwidth]{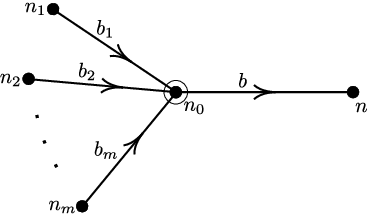}
\end{center}
\caption{\small  Reactants injected at different initial nodes.  }
\label{ExampleNetwork5}
\end{wrapfigure}

A simple variant of the previous example helps to illustrate the situation in which a mixture is injected over several nodes. Consider the reactor of Figure \ref{ExampleNetwork5}. Suppose   $m$ different gas species. A unit pulse containing fractions  $\alpha_1, \dots, \alpha_m$ of each species $1, \dots, m$ is injected so that species $i$ is injected at node $n_i$. Then the fraction of $j$ in the output    is 
$$f_j=\alpha_1f_{1j}(n_1)+\cdots +\alpha_m f_{mj}(n_m).$$
It is easily shown (similarly to the first example) that 
$f(n_1)=\cdots =f(n_m)=f(n_0)$, independently of the lengths and advection velocities of branches $b_1, \dots, b_m$.

The matrix $f(n_0)$ is also easily obtained:
$$ f(n_0)=\left(I-\frac{(m+1)\tilde{\ell}(b)}{D}K\right)^{-1}.$$
So, in fact, $f_j=\alpha_{1}f_{1j}(n_0)+\cdots + \alpha_m f_{mj}(n_0)$. This means that, in this specific situation (where the injection nodes are connected to the exit node through a single path passing through  the one active node $n_0$), the process is equivalent to injecting the whole pulse mixture at once at $n_0$. Incidentally, it is 
not difficult to show directly that a square matrix of the form $I-\mu K$, where $\mu\geq 0$ and $K=(K_{ij})$ is such that $K_{ij}\geq 0$ for $i\neq j$, $K_{ii}=-\sum_{j\neq i}K_{ij}$, is always invertible. For the special case $m=2$ and $ K=\left(\begin{array}{rr}-k_+ & k_+ \\[6pt]k_- & -k_-\end{array}\right),$ we have
$$f(n_0)=\frac1{1+\frac{3\tilde{\ell}(b)}{D}(k_-+k_+)}\left(\begin{array}{cc}1+\frac{3\tilde{\ell}(b)}{D}k_- & \frac{3\tilde{\ell}(b)}{D}k_+ \\[6pt]\frac{3\tilde{\ell}(b)}{D}k_- & 1+\frac{3\tilde{\ell}(b)}{D}k_+\end{array}\right) $$

\subsection{A segment reactor with one active node and a bypass.}
\begin{wrapfigure}{l}{0.3\textwidth}
\begin{center}
 \includegraphics[width=0.28\textwidth]{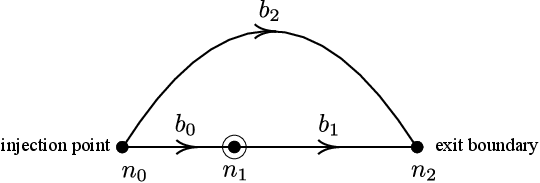}
\end{center}
\caption{\small  A simple network reactor with a single active node and a bypass.  }
\label{ExampleNetwork1}
\end{wrapfigure}

The network reactor for this example is shown in Figure \ref{ExampleNetwork1}.  One easily finds 
$$\eta(n_0,n_1)=\frac{\tilde{\ell}_2}{\tilde{\ell}_0 +\tilde{\ell}_2}, \ \  
\eta(n_1,n_0)=\frac{\tilde{\ell}_1}{\tilde{\ell}_{\overline{0}} +\tilde{\ell}_1}, \ \ \eta(n_0,n_2)=\frac{\tilde{\ell}_0}{\tilde{\ell}_0 +\tilde{\ell}_2},\ \ 
\eta(n_1,n_2)=\frac{\tilde{\ell}_{\overline{0}}}{\tilde{\ell}_{\overline{0}} +\tilde{\ell}_1}$$
so that 
$$\left(\begin{array}{cc}-I & \frac{\tilde{\ell}_2}{\tilde{\ell}_0+\tilde{\ell}_2}I \\[6pt]\frac{\tilde{\ell}_1}{\tilde{\ell}_{\overline{0}}+\tilde{\ell}_1}I & \tilde{K}-I\end{array}\right) 
\left(\begin{array}{c}f(n_0) \\[6pt]f(n_1)\end{array}\right)=-\left(\begin{array}{c}\frac{\tilde{\ell}_0}{\tilde{\ell}_0+\tilde{\ell}_2}I \\[6pt]\frac{\tilde{\ell}_{\overline{0}}}{\tilde{\ell}_{\overline{0}}+\tilde{\ell}_1}I\end{array}\right).$$
 The first equation in this system may be written as
\begin{equation}\label{1equ}f(n_0)=\frac{\tilde{\ell}_2}{\tilde{\ell}_0+\tilde{\ell}_2}f(n_1)+ \frac{\tilde{\ell}_0}{\tilde{\ell}_0+\tilde{\ell}_2}I,\end{equation}
which has a natural interpretation. If we regard the velocity-adjusted length $\tilde{\ell}(b)$ as the resistance to crossing branch $b$, then
$1/\tilde{\ell}(b)$ may be interpreted as a  conductivity.

Transport from $n_0$ to $n_1$ and from $n_0$ to $n_2$ is then distributed in proportion  to the relative conductivity of the two channels. Writing these relative conductivities as $a_1$ and $a_2$ ($a_1+a_2=1$) then Equation
(\ref{1equ}) becomes $$f_{ij}(n_0)=a_1 f_{ij}(n_1)+a_2f_{ij}(n_2)$$ for each $(i,j)$.  Expressed in words, 
the molar fraction of $j$ in the output mixture given that a unit pulse of $i$ is injected at $n_0$ equals the weighted average of these molar fractions for the injection points $n_1$ ($f_{ij}(n_1)$) and $n_2$ ($f_{ij}(n_2)=\delta_{ij}$), with weights given by  the channels' relative conductivity.

It remains to obtain $f(n_1)$,  the output   composition when the  injection is at the active node.  This is easily found to be
\begin{equation}\label{2equ}
f(n_1)=\left(I-\frac{\left(\tilde{\ell}_{\overline{0}}+\tilde{\ell}_1\right)\left(\tilde{\ell}_0+\tilde{\ell}_2\right)}{\tilde{\ell}_{\overline{0}}\tilde{\ell}_0 +\tilde{\ell}_{\overline{0}}\tilde{\ell}_2+\tilde{\ell}_{1}\tilde{\ell}_0}\tilde{K}\right)^{-1}=
\left(I-\alpha {K}\right)^{-1},
\end{equation}
where  
$$\alpha:=\frac{2 \tilde{\ell}_{\overline{0}}\tilde{\ell}_1\left(\tilde{\ell}_0+\tilde{\ell}_2\right)}{D\left(\tilde{\ell}_{\overline{0}}\tilde{\ell}_0 +\tilde{\ell}_{\overline{0}}\tilde{\ell}_2+\tilde{\ell}_{1}\tilde{\ell}_0\right)}, \ \ \tilde{K}:= \frac{2\tilde{\ell}_{\overline{0}} \tilde{\ell}_1}{D\left(\tilde{\ell}_{\overline{0}}+ \tilde{\ell}_1\right)}K. $$
In the special case of only two species, 
$$f(n_1) =  \frac1{1+\alpha(k_-+k_+)}\left(\begin{array}{cc}1+\alpha k_- & \alpha k_+ \\\alpha k_- & 1+\alpha k_+\end{array}\right).$$
 We highlight again the reciprocal relation
 $$\frac{f_{12}(n_0)}{f_{21}(n_0)}=\frac{f_{12}(n_1)}{f_{21}(n_1)}=\frac{k_+}{k_-}.$$

\subsection{The general single active node network}
\begin{figure}[htbp]
\begin{center}
 \includegraphics[width=2.5in]{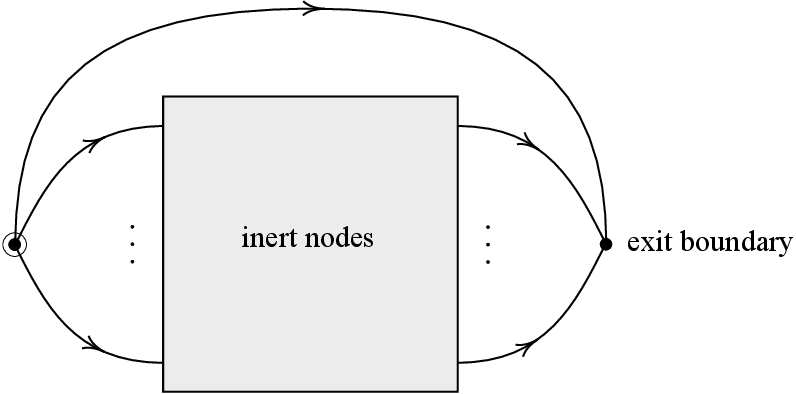}
\caption{\small The general network reactor with a single active node.  The box contains a network of $L$  inert nodes connected to the single active node, $n_0$, on the
left and the exit node, $n_\text{\tiny exit}$, on the right.}
\label{ExampleNetwork2}
\end{center}
\end{figure}

A diagram for this type of reactor is shown in Figure \ref{ExampleNetwork2}. The system for $f$ is 
$$ \left(\begin{array}{cccc}\tilde{K}-I & \eta(n_0,n_1)I & \cdots & \eta(n_0,n_L)I \\[6pt]\eta(n_1,n_0)I & -I & \cdots & \eta(n_1,n_L)I \\[6pt]\vdots & \vdots & \ddots & \vdots \\[6pt]\eta(n_L,n_0)I & \eta(n_L,n_1)I & \cdots & -I\end{array}\right)\left(\begin{array}{c}f(n_0) \\[6pt]f(n_1) \\[6pt]\vdots \\[6pt]f(n_L)\end{array}\right)=
-\left(\begin{array}{c}\eta(n_0,n_{\text{\tiny exit}}) \\[6pt]\eta(n_1,n_{\text{\tiny exit}}) \\[6pt]\vdots \\[6pt]\eta(n_L,n_{\text{\tiny exit}})\end{array}\right)$$
It is implicit in this expression that $\eta(n,n')=0$ when $n$ and $n'$ are not connected by a branch.  Note that all the block entries of the coefficient matrix commute with each other. 

By Cramer's rule for the inverse of the coefficient matrix $\Lambda$, we obtain
\begin{equation}\label{single}f(n_\text{\tiny init})=\left({\alpha(n_\text{\tiny init})I+\beta(n_\text{\tiny init}){K}}\right)\left({\delta(n_\text{\tiny init}) I+\gamma(n_\text{\tiny init}) {K}}\right)^{-1} \end{equation}
where $\alpha, \beta, \gamma, \delta$ are polynomial functions of the velocity-adjusted branch lengths that take into account the geometry and topology of the network reactor as well as the point of injection $n_\text{\tiny init}$. 
From the property that the sum of each row of $f(n_{\text{\tiny init}})$ is $1$ and the sum of each row of $K$ is $0$, it  can be shown     that $\alpha=\delta$. So $f(n_{\text{\tiny init}})$ can be written as
\begin{equation}\label{single_bis}
f(n_{\text{\tiny init}})=(I +\xi(n_{\text{\tiny init}})K)(I+\eta(n_{\text{\tiny init}})K)^{-1},
\end{equation}
where $\xi=\beta/\alpha$ and $\eta=\gamma/\alpha$. From this remark, it is not difficult to conclude that if $c=(c_1, \dots, c_N)$ is the vector of equilibrium concentrations for the reaction matrix $K$ for a closed reactor, so that $cK=0$, then 
 $$cf(n_{\text{\tiny init}}) =c.$$
 This means that if the proportions of the substances in the input composition are the same as for the equilibrium concentrations in a closed reactor, then the output composition is the same as the input composition.

For a concrete example, consider  the reaction system $1\rightleftharpoons 2$ with matrix of reaction coefficients
$$K=\left(\begin{array}{rr}-k_+ & k_+ \\[6pt]k_- & -k_-\end{array}\right), $$
we obtain (using $\alpha =\alpha(n_\text{\tiny init})$, etc.)
\begin{align*}f(n_{\text{\tiny init}})&= \frac{1}{\alpha(\delta -\gamma(k_-+k_+))}  \left(\begin{array}{cc}\alpha \delta -\alpha \gamma k_- -\beta \delta k_+ & (\beta\delta-\alpha \gamma) k_+ \\[6pt](\beta \delta -\alpha \gamma)k_- & \alpha \delta-\alpha\gamma k_+-\beta\delta k_-\end{array}\right) \\[10pt]
&=\frac1{1-\eta(k_++k_-)}
\left(\begin{array}{rr}1-\eta k_--\xi k_+ & (\xi-\eta)k_+ \\[6pt](\xi-\eta)k_- & 1-\xi k_--\eta k_+\end{array}\right)\end{align*}
In particular, for a general network reactor with a single active node and one pair of reversible reactions involving only two substances, the reciprocal relation $f_{12}(n_{\text{\tiny init}})/f_{21}(n_{\text{\tiny init}})=k_+/k_-$ holds.

\subsection{A linear network reactor with two active nodes}
\begin{figure}[htbp]
\begin{center}
 \includegraphics[width=3.0in]{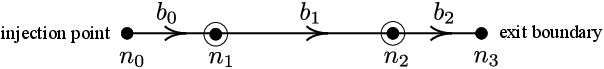}
\caption{\small Linear network with two active nodes.}
\label{ExampleNetwork3}
\end{center}
\end{figure}

 Let us consider the linear network reactor with two active nodes shown in Figure \ref{ExampleNetwork3}.
 
Using the same notational conventions as before, we have the linear system
$$\left(\begin{array}{ccc}-I & I & 0 \\[6pt]\frac{\tilde{\ell}_1}{\tilde{\ell}_{\overline{0}}+\tilde{\ell}_1}I & \tilde{K}_1-I & \frac{\tilde{\ell}_{\overline{0}}}{\tilde{\ell}_{\overline{0}}+\tilde{\ell}_1}I \\[6pt]0 & \frac{\tilde{\ell}_2}{\tilde{\ell}_{\overline{1}}+\tilde{\ell}_2}I & \tilde{K}_2-I\end{array}\right)\left(\begin{array}{c}f(n_0) \\[6pt]f(n_1) \\[6pt]f(n_2)\end{array}\right) =-\left(\begin{array}{c}0 \\[6pt]0 \\[6pt]\frac{\tilde{\ell}_{\overline{1}}}{\tilde{\ell}_{\overline{1}}+\tilde{\ell}_2}I\end{array}\right)$$
The solution  may be written in matrix form as
\begin{align*}
f(n_2)&= \left[I -\frac{2}{D}\left(
\frac{\tilde{\ell}_1 \left(\tilde{\ell}_{\overline{1}}+\tilde{\ell}_2\right)}{\tilde{\ell}_{\overline{1}}} K_1 +\tilde{\ell}_2 K_2
\right)
+\frac{4}{D^2}\tilde{\ell}_1\tilde{\ell}_2 K_1K_2
\right]^{-1}
\left(I-\frac{2\tilde{\ell}_1}{D}K_1\right)\\
f(n_1)&=\left(I-\frac{2\tilde{\ell}_1}{D} K_1\right)^{-1} f(n_2)\\
f(n_0)&= f(n_1).
\end{align*}
In particular,
\begin{equation}\label{fn0}f(n_0) = \left(I-\frac{2\tilde{\ell}_1}{D} K_1\right)^{-1}\left[I -\frac{2}{D}\left(
\frac{\tilde{\ell}_1 \left(\tilde{\ell}_{\overline{1}}+\tilde{\ell}_2\right)}{\tilde{\ell}_{\overline{1}}} K_1 +\tilde{\ell}_2 K_2
\right)
+\frac{4}{D^2}\tilde{\ell}_1\tilde{\ell}_2 K_1K_2
\right]^{-1}
\left(I-\frac{2\tilde{\ell}_1}{D}K_1\right).\end{equation}
Note that the first and third factors in the above expression for $f(n_0)$ cancel out when $K_1$ and $K_2$ commute. This is the case, for example, when 
 the two nodes implement the same reactions, that is $K_1=K_2$, or when they implement reactions involving non-intersecting sets of species, making the reactions at $n_1$ and $n_2$ independent. In the latter case,
 $K_1$ and $K_2$ have diagonal block form 
 $$K_1=\left(\begin{array}{c|c}L_1 & \ \\\hline \ & O_2\end{array}\right),\ \ K_2=\left(\begin{array}{c|c}O_1 & \ \\\hline \ & L_2\end{array}\right),$$
 where $O_1, O_2$ are the zero square matrices, $L_1, L_2$ are rate constant matrices, and $O_i$ and $L_i$ have the same size for $i=1,2$. 
From this form it also follows that  $K_1K_2=0$. Therefore, in this case, we have
$$f(n_0) = \left[I -\frac2D   \left(\begin{array}{c|c}\frac{\tilde{\ell}_1 \left(\tilde{\ell}_{\overline{1}}+\tilde{\ell}_2\right)}{\tilde{\ell}_{\overline{1}}} L_1 & \ \\[6pt]\hline \ & \tilde{\ell}_2L_2\end{array}\right)    \right]^{-1}. $$ 
 
 Let us write the solution more explicitly in a couple of special cases. First suppose that  the only reactions are
 $1\rightleftharpoons 2$, and they have the same constants at $n_1$ and $n_2$. That is,
 $$K_1=K_2 =\left(\begin{array}{rr}-k_+ & k_+ \\[6pt]k_- & -k_-\end{array}\right). $$
 Then
  $$f(n_0) = \frac1{1+\alpha(k_-+k_+) +\beta (k_-+k_+)^2}
   \left(\begin{array}{cc}  1+\alpha k_- + \beta k_-(k_-+k_+) & \alpha k_+ + \beta k_+(k_-+k_+) \\[6pt] \alpha k_- + \beta k_-(k_-+k_+) &
    1+\alpha k_+ + \beta k_+(k_-+k_+)   \end{array}\right)$$
 where 
 $$\alpha= \frac2D\frac{\tilde{\ell}_1\tilde{\ell}_2+\tilde{\ell}_1\tilde{\ell}_{\overline{1}}+ \tilde{\ell}_2\tilde{\ell}_{\overline{1}}}{\tilde{\ell}_1} , \ \  \beta=\frac{4}{D^2}\tilde{\ell}_1\tilde{\ell}_2.$$
 Thus, for example, if $2$ is initially injected at $n_0$, the molar fraction of $1$ in the output composition is
 $$f_{21}(n_0)=  \frac{ \alpha k_- + \beta k_-(k_-+k_+)}{1+\alpha(k_-+k_+) +\beta (k_-+k_+)^2}. $$

Here again, as in the first example, we observe the symmetric relation:
$$\frac{f_{12}(n_0)}{f_{21}(n_0)}=\frac{\alpha k_+ +\beta k_+ (k_-+k_+)}{\alpha k_- +\beta k_- (k_-+k_+)}=\frac{k_+}{k_-}.$$

As a second instance of the same network, let us suppose that there are $3$ chemical species, and that $n_1$ implements the reaction $1\rightleftharpoons 2$ and $n_2$ implements the reaction $2\rightleftharpoons 3$. The reaction matrices for this system are
$$K_1= \left(\begin{array}{ccc}\! -k_+ & k_+ & 0 \\[6pt] \ k_- &\! \! -k_- & 0 \\[6pt]0 & 0 & 0\end{array}\right),\ \ K_2= \left(\begin{array}{ccc}0 & 0 & 0 \\[6pt]0 &\! -k_+' & k_+' \\[6pt]0 &\  k_-' &\! \! -k_-'\end{array}\right).$$
Let us define the constants $$\alpha = \frac{2}{D}\tilde{\ell}_1, \ \  \beta=\frac{2}{D}\frac{\tilde{\ell}_1\left(\tilde{\ell}_2 +\tilde{\ell}_{\overline{1}}\right)}{\tilde{\ell}_{\overline{1}}}, \ \ \gamma=\frac2D\tilde{\ell}_2, \ \ \delta=\frac4{D^2} \tilde{\ell}_1\tilde{\ell}_2.$$
Then, using Equation (\ref{fn0}),
$$f(n_0)=\left(\begin{array}{ccc}1+\alpha k_+ & -\alpha k_+ & 0 \\[6pt]-\alpha k_- & 1+\alpha k_- & 0 \\[6pt]0 & 0 & 1\end{array}\right)^{-1}   \left(\begin{array}{ccc}1+\beta k_+ & -\beta k_+ + \delta k_+ k_+' & \delta k_+ k_+' \\[6pt]-\beta k_- & 1+\beta k_- +\gamma k_+' + \delta k_-k_+' & -\gamma k_+'-\delta k_-k_+' \\[6pt]0 & -\gamma k_-' & 1+\gamma k_-'\end{array}\right)^{-1} \left(\begin{array}{ccc}1+\alpha k_+ & -\alpha k_+ & 0 \\[6pt]-\alpha k_- & 1+\alpha k_- & 0 \\[6pt]0 & 0 & 1\end{array}\right).$$
From the evaluation of this matrix, we can find, for example, $f_{31}(n_0)$, the fraction represented by species $1$ in the output given an initial unit pulse of $3$ injected at $n_0$:
$$f_{31}(n_0)=\frac{\delta (\tilde{\ell}_2/\tilde{\ell}_{\overline{1}}){k_-k_-'}}{1+\beta(k_-+k_+)+\gamma(k_-'+k_+')+\beta \gamma (k_-k_-'+k_-'k_++k_+k_+')+\delta k_-k_+'+2\beta\delta k_-k_+k_+'+2\beta\gamma\delta k_-k_-'k_+k_+'}.$$
 
\subsection{A network reactor with parallel active nodes}
\begin{figure}[htbp]
\begin{center}
 \includegraphics[width=2.5in]{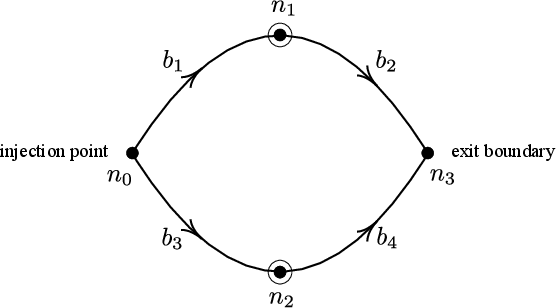}
\caption{\small Two active nodes in parallel. }
\label{ExampleNetwork4}
\end{center}
\end{figure}

To the network of   Figure \ref{ExampleNetwork4} is associated the system
$$\left(\begin{array}{ccc}-I & \eta(n_0,n_1)I & \eta(n_0,n_2)I \\[6pt]\eta(n_1,n_0)I & \tilde{K}_1-I & 0 \\[6pt]\eta(n_2,n_0)I & 0 & \tilde{K}_2-I\end{array}\right) \left(\begin{array}{c}f(n_0) \\[6pt]f(n_1) \\[6pt]f(n_2)\end{array}\right)=-\left(\begin{array}{c}0 \\[6pt]\eta(n_1,n_3)I \\[6pt]\eta(n_2,n_3)I\end{array}\right)$$
where
$$\eta(n_0,n_1)=1-\eta(n_0,n_2)=\frac{\tilde{\ell}_3}{\tilde{\ell}_1+\tilde{\ell}_3}, \ \  
\eta(n_1,n_0)=1-\eta(n_1,n_3)=\frac{\tilde{\ell}_2}{\tilde{\ell}_2+\tilde{\ell}_{\overline{1}}}, \ \
\eta(n_2,n_0)=1-\eta(n_2,n_3)=\frac{\tilde{\ell}_4}{\tilde{\ell}_4+\tilde{\ell}_{\overline{3}}}$$
and 
$$\tilde{K}_1=\frac2D \frac{\tilde{\ell}_2\tilde{\ell}_{\overline{1}}}{\tilde{\ell}_2+\tilde{\ell}_{\overline{1}}}K_1, \ \  
\tilde{K}_2=\frac2D \frac{\tilde{\ell}_4\tilde{\ell}_{\overline{3}}}{\tilde{\ell}_4+\tilde{\ell}_{\overline{3}}}K_2.$$
First note that  
$$f(n_0)= \frac{\tilde{\ell}_3}{\tilde{\ell}_1+\tilde{\ell}_3}f(n_1)+ \frac{\tilde{\ell}_1}{\tilde{\ell}_1+\tilde{\ell}_3}f(n_2).$$
That is, the output composition matrix for the injection point $n_0$ is the weighted average of those with injection points at the active nodes. (The longer the velocity-adjusted length of a branch, the lesser the weight of that branch among the possible channels leading to active nodes.)

To obtain the latter, it helps to set the following notation.
$$a_1=\frac2D\frac{\tilde{\ell}_{\overline{1}}\tilde{\ell}_2(\tilde{\ell}_1+\tilde{\ell}_{3})}{\tilde{\ell}_1\tilde{\ell}_{\overline{1}} +\tilde{\ell}_{\overline{1}}\tilde{\ell}_3+\tilde{\ell}_1\tilde{\ell}_2}, \ \ 
    a_2= \frac{\tilde{\ell}_1\tilde{\ell}_3(\tilde{\ell}_{\overline{1}}+\tilde{\ell}_2)}{(\tilde{\ell}_1+\tilde{\ell}_3)(\tilde{\ell}_1\tilde{\ell}_{\overline{1}} +\tilde{\ell}_{\overline{1}}\tilde{\ell}_3+\tilde{\ell}_1\tilde{\ell}_2)},\ \ 
 a_3=\frac{\tilde{\ell}_3\tilde{\ell}_4}{\tilde{\ell}_3\tilde{\ell}_{4}+\tilde{\ell}_1\tilde{\ell}_{\overline{3}}+\tilde{\ell}_3\tilde{\ell}_{\overline{3}}},\ \ 
 a_4=\frac2D\frac{\tilde{\ell}_{\overline{3}}\tilde{\ell}_4(\tilde{\ell}_1+\tilde{\ell}_3)}{ \tilde{\ell}_3\tilde{\ell}_{4}+\tilde{\ell}_1\tilde{\ell}_{\overline{3}}+\tilde{\ell}_3\tilde{\ell}_{\overline{3}}}   $$
 and
 $$b_1=\frac{\tilde{\ell}_{\overline{1}}(\tilde{\ell}_1+\tilde{\ell}_3)}{\tilde{\ell}_1\tilde{\ell}_{\overline{1}} +\tilde{\ell}_{\overline{1}}\tilde{\ell}_3+\tilde{\ell}_1\tilde{\ell}_2}, \ \ 
 b_2= \frac{\tilde{\ell}_{\overline{3}}(\tilde{\ell}_1+\tilde{\ell}_3)}{\tilde{\ell}_3\tilde{\ell}_{4}+\tilde{\ell}_1\tilde{\ell}_{\overline{3}}+\tilde{\ell}_3\tilde{\ell}_{\overline{3}}}.$$
Then  the matrices $f(n_1), f(n_2)$ satisfy the system
$$(I-a_1K_1)f(n_1)-a_2 f(n_2)=b_1, \ \ -a_3f(n_1)+(I-a_4K_2)f(n_2)=b_2. $$
Among the many possibilities one can explore, let us only write the solution for $K_1=K_2=K$, $\ell_i=\ell$ and $\nu(b_i)=0$  for $i=1,2,3,4$. Then
$\tilde{\ell}_i=\tilde{\ell}_{\overline{i}}=\ell$.
$$\left(\begin{array}{c}f(n_1) \\[6pt]f(n_2)\end{array}\right)=\left[\left(I-\frac{4\ell}{3D}K\right)^2-\frac19I\right]^{-1} \left(\begin{array}{cc}I-\frac{4\ell}{3D}K & \frac13I \\[6pt]\frac13 I & I-\frac{4\ell}{3D}K\end{array}\right) \left(\begin{array}{c}\frac23 \\[6pt]\frac23\end{array}\right). $$
This is easily solved:
$$f(n_1)=f(n_2)= \left(I-\frac{2\ell}{D}K\right)^{-1}. $$
For the simple reaction $1\rightleftharpoons 2$, for which $K=\left(\begin{array}{cc}-k_+ & k_+ \\[6pt]k_- & -k_-\end{array}\right)$,  
$$f(n_1)=f(n_2)=\frac{1}{1+\frac{2\ell}{D}(k_-+k_+)}  \left(\begin{array}{cc}1+\frac{2\ell}{D}k_- & \frac{2\ell}{D}k_+ \\[6pt]\frac{2\ell}{D}k_- & 1+\frac{2\ell}{D}k_+\end{array}\right).$$
 Once again we see the reciprocity relation $$
 \frac{f_{12}(n_0)}{f_{21}(n_0)}=\frac{f_{12}(n_1)}{f_{21}(n_1)}=\frac{f_{12}(n_2)}{f_{21}(n_2)}=\frac{k_+}{k_-}.$$

\subsection{Adsorption}

In this paper we do not yet develop systematically reaction mechanisms  involving adsorption at  active regions, but we illustrate   how such reactions may be handled
based on the methods considered so far. By adsorption we have in mind reactions of the form $A+Z\rightleftharpoons AZ$, where
$Z$ is a catalyst that remains, together with the complex $AZ$, at the active region while  $A$   migrates freely through the reactor via diffusion and advection.

Under the assumption that the amount  and characteristics of   $Z$ at a node remain essentially unchanged over the time span of the process, we may replace the second order reaction with the first order $A\rightleftharpoons AZ$. It is interesting then to investigate whether introducing $AZ$ as a new species having zero advection velocity and very small diffusivity $D$, and then passing to the limit as $D$ approaches $0$, gives meaningful results. We explore this possibility   with one example and leave a systematic treatment  for another study.

\begin{wrapfigure}{r}{0.4\textwidth}
\begin{center}
 \includegraphics[width=0.38\textwidth]{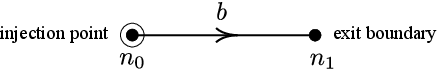}
\end{center}
\caption{\small  A simple network reactor to illustrate adsorption reactions.  }
\label{ExampleNetwork6}
\end{wrapfigure} 

Let us consider the system of reactions
  $$A\rightleftharpoons AZ\rightleftharpoons BZ\rightleftharpoons B$$
where   $A$ and $B$ are gases, subject to adsorption and disorption,   and  $AZ$ and $BZ$ are catalytic intermediates. $A$ and $B$ will be called {\em migrating} and $AZ$ and $BZ$ {\em static}. For simplicity of notation, we use $1, 2, 3, 4$ for $A, AZ, BZ, B$, respectively.  For the reactor we take the (simplest) example shown in Figure \ref{ExampleNetwork6}. See \cite{CYMGI} for a justification of this mechanism.

We suppose that $1$ and $4$ have diffusivity $D$ and advection velocity $\nu=\nu(b)$ while $2$ and $3$ have (small) diffusivity $D'$, which will be taken to $0$ at the end of the calculation, and zero advection velocity. 
Let us indicated the length of $b$ by $\ell$, its velocity-adjusted length by $\tilde{\ell}$ and the  reaction coefficients by
$$
\ce{
1
<=>[\ce{k_+^{(1)}}][\ce{k_-^{(1)}}]
{\ce{2}}
<=>[\ce{k_+}][\ce{k_-}]
{\ce{3}}
<=>[\ce{k_+^{(2)}}][\ce{k_-^{(2)}}]
{\ce{4}}
}.
$$
The output composition matrix is then the $4\times 4$-matrix
$$f(n_0)=\left(I-\tilde{K}\right)^{-1} $$
where
$$\tilde{K}_{ij}=\begin{cases}
 \frac{\tilde{\ell}}{D} K_{ij} & \text{ if } i=1, 4\\[6pt]
 \frac{{\ell}}{D'} K_{ij} & \text{ if } i=2,3
\end{cases} \ \ \text{ or }\ \  \tilde{K}=\left(\begin{array}{cccc}-\frac{\tilde{\ell}}{D}k_+^{(1)} & \frac{\tilde{\ell}}{D}k_+^{(1)} & 0 & 0 \\[6pt]\frac{\ell}{D'}k^{(1)}_- & -\frac{\ell}{D'}\left(k^{(1)}_-+k_+\right) & \frac{\ell}{D'}k_+ & 0 \\[6pt]0 & \frac{\ell}{D'}k_- & -\frac{\ell}{D'}\left(k_-+k_+^{(2)}\right) & \frac{\ell}{D'}k^{(2)}_+ \\[6pt]0 & 0 & \frac{\tilde{\ell}}{D}k_-^{(2)} & -\frac{\tilde{\ell}}{D}k_-^{(2)}\end{array}\right).$$
In the limit as $D'$ approaches $0$ we obtain
\begin{equation}\label{fn0} f(n_0) =\left(\begin{array}{cccc}
\frac{k_-k_-^{(1)}\left(1+\frac{\tilde{\ell}}{D}k_-^{(2)}\right) +k_+^{(2)}\left(k_++k_-^{(1)}\right)}{k_+^{(2)}\left[k_-^{(1)}+\left(1+\frac{\tilde{\ell}}{D}k_+^{(1)}\right)k_+\right] +k_-k_-^{(1)}\left(1+\frac{\tilde{\ell}}{D}k_-^{(2)}\right)}& 
0 & 0 & \frac{\frac{\tilde{\ell}}{D}k_+k_+^{(1)}k_+^{(2)}}{{k_+^{(2)}\left[k_-^{(1)}+\left(1+\frac{\tilde{\ell}}{D}k_+^{(1)}\right)k_+\right] +k_-k_-^{(1)}\left(1+\frac{\tilde{\ell}}{D}k_-^{(2)}\right)}}\\[10pt]
\frac{k_-^{(1)}\left[k_+^{(2)}+k_-\left(1+\frac{\tilde{\ell}}{D}k_-^{(2)}\right)\right]}{{k_+^{(2)}\left[k_-^{(1)}+\left(1+\frac{\tilde{\ell}}{D}k_+^{(1)}\right)k_+\right] +k_-k_-^{(1)}\left(1+\frac{\tilde{\ell}}{D}k_-^{(2)}\right)}} & 
0 & 0 & \frac{k_-k_+\left(1+\frac{\tilde{\ell}}{D}k_+^{(1)}\right)}{{k_+^{(2)}\left[k_-^{(1)}+\left(1+\frac{\tilde{\ell}}{D}k_+^{(1)}\right)k_+\right] +k_-k_-^{(1)}\left(1+\frac{\tilde{\ell}}{D}k_-^{(2)}\right)}} \\[10pt]
\frac{k_-k_-^{(1)}\left(1+\frac{\tilde{\ell}}{D}k_-^{(2)}\right)}{{k_+^{(2)}\left[k_-^{(1)}+\left(1+\frac{\tilde{\ell}}{D}k_+^{(1)}\right)k_+\right] +k_-k_-^{(1)}\left(1+\frac{\tilde{\ell}}{D}k_-^{(2)}\right)}} & 
0 & 0 & \frac{k_+^{(2)}\left[k_-^{(1)}+k_+\left(1+\frac{\tilde{\ell}}{D}k_+^{(1)}\right)\right]}{{k_+^{(2)}\left[k_-^{(1)}+\left(1+\frac{\tilde{\ell}}{D}k_+^{(1)}\right)k_+\right] +k_-k_-^{(1)}\left(1+\frac{\tilde{\ell}}{D}k_-^{(2)}\right)}}\\[10pt]
\frac{\frac{\tilde{\ell}}{D}k_-k_-^{(1)}k_-^{(2)}}{{k_+^{(2)}\left[k_-^{(1)}+\left(1+\frac{\tilde{\ell}}{D}k_+^{(1)}\right)k_+\right] +k_-k_-^{(1)}\left(1+\frac{\tilde{\ell}}{D}k_-^{(2)}\right)}} & 
0 & 0 &\frac{k_-k_-^{(1)}+k_+^{(2)}\left[k_-^{(1)}+k_+\left(1+\frac{\tilde{\ell}}{D}k_+^{(1)}\right)\right]}{{k_+^{(2)}\left[k_-^{(1)}+\left(1+\frac{\tilde{\ell}}{D}k_+^{(1)}\right)k_+\right] +k_-k_-^{(1)}\left(1+\frac{\tilde{\ell}}{D}k_-^{(2)}\right)}}\end{array}\right). \end{equation}
As was to be expected, $f_{ij}(n_0)=0$ whenever $j=2, 3$. This is because the species $AZ$ and $BZ$ should remain at $n_0$ and not migrate to the exit. It is interesting to compare
the fraction of $B$ in the output given a unit pulse of $A$ injected at $n_0$ and the corresponding value for the overall reaction $A\rightleftharpoons B$. The former is 
$$f_{14}(n_0)= \frac{\frac{\tilde{\ell}}{D}k_+k_+^{(1)}k_+^{(2)}}{{k_+^{(2)}\left[k_-^{(1)}+\left(1+\frac{\tilde{\ell}}{D}k_+^{(1)}\right)k_+\right] +k_-k_-^{(1)}\left(1+\frac{\tilde{\ell}}{D}k_-^{(2)}\right)}}.$$
For the latter, let us indicate the reaction coefficients by $K_\pm$. Then the
fraction of $B$ in the output given that a unit pulse of $A$ was injected at $n_0$ is
$$f_{AB}(n_0)= \frac{\frac{\tilde{\ell}}{D}K_+}{1+\frac{\tilde{\ell}}{D}\left(K_-+K_+\right)}.$$
Note that, if $k_-^{(1)}=k_-^{(2)}=k_+^{(2)}=k_+^{(1)}$ is taken to be very large so that  adsorption/desorption  are very fast reactions,  then $f_{14}(n_0)$ reduces to $f_{AB}(n_0)$ for $K_\pm=k_\pm$. 

Finally, it is again notable from Equation (\ref{fn0}) the reciprocal relation
$$\frac{f_{14}(n_0)}{f_{41}(n_0)}=\frac{k_+ k^{(1)}_+ k^{(2)}_+}{k_- k^{(1)}_- k^{(2)}_-},$$
which does not depend on transport coefficients. 

\subsection{Numerical investigation of the network of Figure \ref{ExampleNetwork}}
Examining the system of equations (\ref{big}), we immediately see that 
  $f(n_0)=f(n_1)$. This means that injecting the initial pulse at $n_0$ has exactly the same effect as injecting it at $n_1$. By discarding the branch $b_0$, we obtain a simpler system 
$$\left(\begin{array}{cccc} -I & \eta(n_1,n_2)I & \eta(n_1,n_3)I & \eta(n_1,n_4)I \\[6pt] \eta(n_2,n_1)I & \tilde{K}(n_2)-I & 0 & \eta(n_2,n_4)I \\[6pt] \eta(n_3,n_1)I & 0 & \tilde{K}(n_3)-I & \eta(n_3,n_4)I \\[6pt] \eta(n_4,n_1)I & \eta(n_4,n_2)I & \eta(n_4,n_3)I & -I\end{array}\right) \left(\begin{array}{c}f(n_1) \\[6pt]f(n_2) \\[6pt]f(n_3) \\[6pt]f(n_4)\end{array}\right)=-\left(\begin{array}{c}  0 \\[6pt] 0 \\[6pt] 0\\[6pt]\eta(n_4,n_5)\end{array}\right).$$

\begin{figure}[htbp]
\begin{center}
 \includegraphics[width=6.0in]{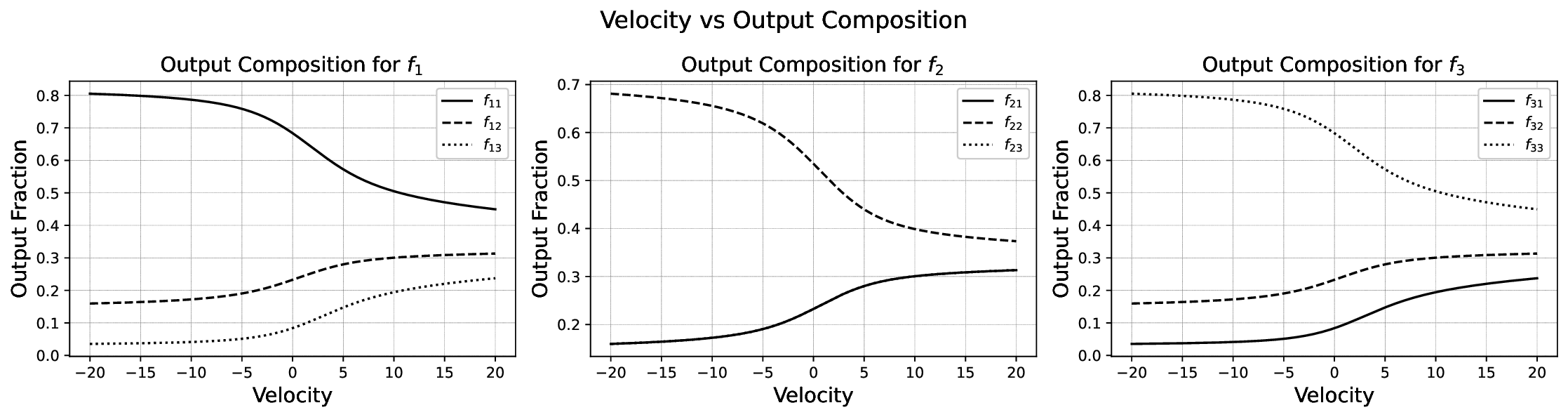}
\caption{\small Dependence of the $f_{ij}(n_0)$ on the advection velocity on $b_5$ for the network example of Figure \ref{ExampleNetwork}. In the middle plot, the curves for $f_{21}$ and $f_{23}$ agree, so the latter is not apparent.} 
\label{plots2}
\end{center}
\end{figure} 
It makes sense, conceptually, and for the purpose of simplifying expressions, to introduce $h_i:=\frac1{\tilde{\ell}(\mathbf{b}_i)}$, $h_{\overline{i}}=\frac1{\tilde{\ell}(\overline{\mathbf{b}}_i)}$. 
We may think of this reciprocal of the velocity-adjusted length as a measure of conductance: the larger $h$   the easier it is to move along the corresponding   $\mathbf{b}$.

\begin{figure}[htbp]
\begin{center}
 \includegraphics[width=6.0in]{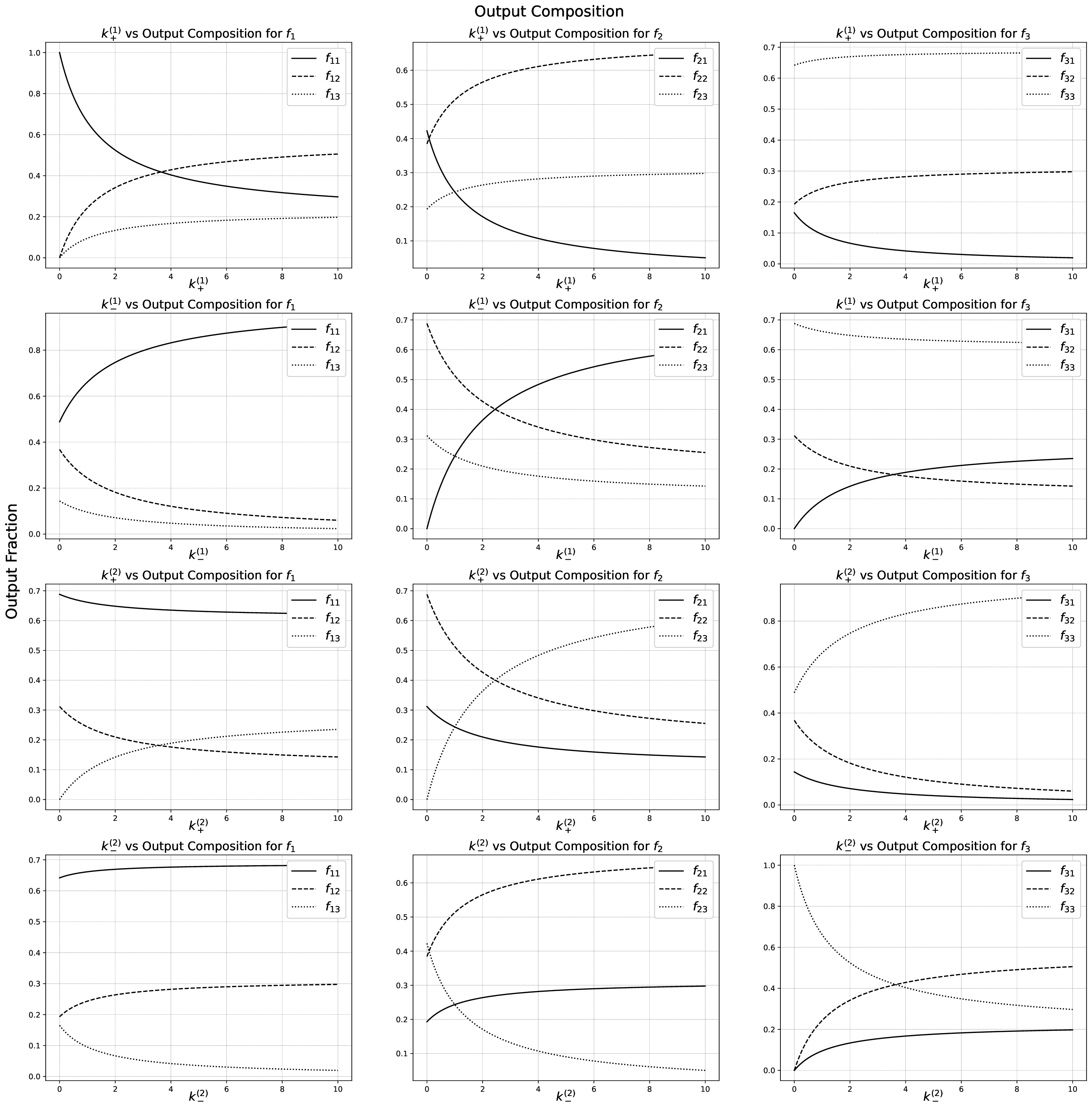}
\caption{\small  Dependence of the $f_{ij}$ on the reaction coefficients $k^{1}_\pm$ and $k^{(2)}_\pm$ for the network example of Figure \ref{ExampleNetwork}.} 
\label{plots1}
\end{center}
\end{figure} 

 Written in terms of conductances, the above system becomes
$$\left(\begin{array}{cccc} -I & \frac{h_1}{h_1+h_3+h_{\overline{5}}}I & \frac{h_3}{h_1+h_3+h_{\overline{5}}}I & \frac{h_{\overline{5}}}{h_1+h_3+h_{\overline{5}}}I \\[6pt] 
\frac{h_{\overline{1}}}{h_{\overline{1}}+h_2}I & \frac{2{K}(n_2)}{D(h_{\overline{1}}+h_2)}-I & 0 &\frac{h_2}{h_{\overline{1}}+h_2}I \\[6pt] 
\frac{h_{\overline{3}}}{h_{\overline{3}}+h_4}I & 0 & \frac{2{K}(n_3)}{D(h_{\overline{3}}+h_4)}-I & \frac{h_4}{h_{\overline{3}}+h_4}I \\[6pt] 
\frac{h_5}{h_{\overline{2}}+h_{\overline{4}}+h_5+h_6}I &\frac{h_{\overline{2}}}{h_{\overline{2}}+h_{\overline{4}}+h_5+h_6}I & \frac{h_{\overline{4}}}{h_{\overline{2}}+h_{\overline{4}}+h_5+h_6}I & -I\end{array}\right) 
\left(\begin{array}{c}f(n_1) \\[6pt]f(n_2) \\[6pt]f(n_3) \\[6pt]f(n_4)\end{array}\right)=-\left(\begin{array}{c}  0 \\[6pt] 0 \\[6pt] 0\\[6pt]
\frac{h_6}{h_{\overline{2}}+h_{\overline{4}}+h_5+h_6}\end{array}\right).$$
In order to keep the number of parameters reasonably small, let us suppose that all branch lengths $\ell(b)=\ell$ are equal and that the only non-zero advection velocity is on $b_5$.  (See Figure \ref{ExampleNetwork}.) Let us further choose the following reactions at nodes $n_2$ and $n_3$:
$$
\ce{1
<=>[k^{(1)}_+][k^{(1)}_-]{2}
}  \text{ at }  n_2, \ \ \ \ 
\ce{2
<=>[k^{(2)}_+][k^{(2)}_-]{3}
}  \text{ at }  n_3.
$$
We thus have $5$ parameters (keeping the value of $\ell$ fixed), which can be written in dimensionless form:
$$s:=\frac{\ell \nu(\mathbf{b}_5)}{D}, \ \ \tilde{k}^{(1)}_\pm :=\frac{\ell k^{(1)}_\pm}{D},\ \   \tilde{k}^{(2)}_\pm :=\frac{\ell k^{(2)}_\pm}{D}.$$
The  pairs  $\tilde{k}^{(1)}_\pm $ and $\tilde{k}^{(2)}_\pm $ 
are a type  of Damkoehler numbers.

The velocity-adjusted length of $\mathbf{b}_5$ is then a function of $s$:
$$\tilde{\ell}(\mathbf{b}_5) =\begin{cases} \frac{1-e^{-s}}{s}\ell &\text{ if } s>0\\[6pt]
\frac{e^{|s|}-1}{|s|}\ell & \text{ if } s<0.
\end{cases}$$
For very large $|\nu(\mathbf{b}_5)|$, if the velocity is positive,  the adjusted length is very short and the nodes $n_1$ and $n_2$ effectively collapse into one; if the velocity is negative, the adjusted length is very long and this middle branch connecting $n_1$ and $n_2$ is effectively removed.

The plots in Figures \ref{plots2} and \ref{plots1} show how the fractions $f_{ij}(n_0)$ vary as functions of the above $5$ parameters. We have fixed $\ell=1$ and $D=1$ so that $s=\nu(\mathbf{b}_5)$ and $\tilde{\kappa}^{(i)}_\pm={\kappa}^{(i)}_\pm.$ It is clear from the graphs that $\sum_jf_{ij}(n_0)=1$.

\section{The theory behind the solution to the OCP} \label{proof}
\subsection{The reaction-transport equations and their fundamental solution}\label{proof_1}
We consider the reaction-transport equations
$$ \frac{\partial c_i}{\partial t}+\nabla\cdot\mathbf{j}_i=\sum_jc_j\kappa_{ji},$$
where $\mathbf{j}_i=c_i\mathbf{v}_i -D_i\nabla c_i$ is the flux vector field of species $i=1, \dots, N$, subject to the boundary conditions:  
\begin{itemize}
\item $c_i(x,t)=0$ for $x$ on the exit boundary of $\mathcal{R}$;
\item the normal component of the flux, $\mathbf{n}(x)\cdot \mathbf{j}_i(x,t)$, equals zero on the reflecting boundary.
\end{itemize}  Setting up the notation
$$(\mathcal{L}c)_i=\nabla\cdot\left(D_i\nabla c_i-c_i\mathbf{v}_i\right)+\sum_jc_j\kappa_{ji},$$
we have the system of partial differential equations $\frac{\partial c}{\partial t}=\mathcal{L}c$, where $c=(c_i)$ is regarded as a vector-valued function of $x$ and $t$. For a  function $\phi(x)$, not necessarily representing  a  concentration, it will occasionally be useful to write $\mathbf{j}_{i,\phi}:=\phi\mathbf{v}_i - D_i\nabla\phi$, the {\em flux} associated to  $\phi$ and $i$.
This reaction-transport model is a special case of the more general setting presented in \cite{GSW}.

The {\em fundamental solution} to the reaction-transport system of equations is the function $\rho_{ij}(x,t|y,s)$ $(t>s)$ representing the concentration of $i$ at position and time $(x,t)$ given that a unit pulse of $j$ is injected into $\mathcal{R}$ at position and time $(y,s)$. This means that 
$$\frac{\partial \rho_{ij}}{\partial t} = \nabla\cdot\left(D_i\nabla \rho_{ij} -\rho_{ij}\mathbf{v}_i\right) +\sum_k\rho_{kj}\kappa_{ki} $$
holds on $\mathcal{R}$ and    
\begin{itemize}
\item $\rho_{ij}(x,t|y,s)=0$ for $x$ on the exit boundary $\partial_{\text{\tiny exit}}\mathcal{R}$;
\item $\mathbf{n}(x)\cdot \left(D_i(x)\nabla_x \rho_{ij}(x,t|y,s) -\rho_{ij}(x,t|y,s)\mathbf{v}_i(x)\right)=0$  for $x$ on the reflecting boundary $\partial_{\text{\tiny reflect}}\mathcal{R}$;
\item $\int_{\mathcal{R}}\rho(x,t|y,s)\phi(y)\, dy\rightarrow \phi(x)$ as $t$ approaches $s$ from above.
\end{itemize}
The integrand in the last item should be interpreted as matrix product: 
$$\sum_j\int_{\mathcal{R}} \rho_{ij}(x,t|y,s)\phi_j(y)\, dy\rightarrow \phi_i(x).$$
This means that $$\rho_{ij}(x,t|y,s)\rightarrow \delta(x-y)\delta_{ij} $$
where $\delta(x)$ is Dirac's delta supported at $0$ and $\delta_{ij}$ is Kronecker's delta. 
When the coefficients of the reaction-transport equation are independent of $t$, we have $\rho(x,t|y,s)=\rho(x,t-s|y,0)$.

It can be shown (by the uniqueness of the fundamental solution) that $\rho$ satisfies the Chapman-Kolmogorov equation
$$ \rho(x,t|y,s)=\int_\mathcal{R} \rho(x,t|\xi,\tau)\rho(\xi,\tau |y,s)\, d\xi$$
where the integrand involves matrix multiplication and $s<\tau<t$.  
This relation can be expressed in operator form as follows. Define for $t>0$ and a row vector-valued function $\phi(y)=(\phi_i(y))$
$$(P_t\phi)(x):=\int_\mathcal{R} \phi(y)\rho(y,t|x,0)\, dy. $$
Note that $(\phi\rho)_i=\sum_j\phi_j\rho_{ji}$. Using the Chapman-Kolmogorov equation it is not difficult to obtain the semigroup property $$P_{t_1+t_2}=P_{t_1}\circ P_{t_2}.$$
To the operator semigroup is associated its generator, which is the operator  $\mathcal{A}$ defined by
$$\mathcal{A}\phi :=\lim_{t\rightarrow 0}\frac{P_t\phi -\phi}{t}$$
on a domain consisting of the functions $\phi$ for which the limit exists.

\subsection{Relation between $\mathcal{A}$ and $\mathcal{L}$}\label{relation}
We wish to characterize $\mathcal{A}$ as the (Hilbert space) adjoint of $\mathcal{L}$. For this, consider the Hilbert space of square-integrable vector-valued (complex) functions on $\mathcal{R}$ with the inner product
$$\langle \varphi,\psi\rangle=\sum_j\int_{\mathcal{R}}\overline{\varphi_j(x)}\psi_j(x)\, dx.$$
The differential operator $\mathcal{L}$ is more precisely defined on the dense  subspace of this Hilbert space consisting of continuous functions $\varphi=(\varphi_i)$ such that each $\varphi_i$ is continuous and has square-integrable (weak) derivatives up to second order. In addition, functions in the domain of $\mathcal{L}$ are zero on the exit boundary of $\mathcal{R}$ and have $0$ normal flux component, $\mathbf{n}(x)\cdot\mathbf{j}_{i,\varphi_i}(x)=0$, at
all $x$ on the reflecting boundary.

A standard integration by parts computation using the divergence theorem gives:
\begin{align*}
\langle \varphi, \mathcal{L}\psi\rangle&=\sum_{j}\int_{\mathcal{R}}\overline{[\nabla\cdot(D_j\nabla\varphi_i)+\mathbf{v}_j\cdot \nabla\varphi_j+\sum_k\kappa_{jk}\varphi_k]}\psi_j\, dx\\[6pt]
&\ \ \ \ \ \ \ \ \ \ \ \ \ \sum_j\left(\int_{\partial_{\text{\tiny exit}}\mathcal{R}} \overline{\varphi}_j D_j \mathbf{n}\cdot\nabla\psi_j\, dA   + \int_{\partial_{\text{\tiny reflect}} \mathcal{R}} \psi_j D_j \mathbf{n}\cdot\nabla\overline{\varphi}_j\, dA\right).
\end{align*}
For $\langle \varphi, \mathcal{L}\psi\rangle$ to be  a bounded linear functional on the domain of $\mathcal{L}$ it is necessary and sufficient that 
$\varphi_j$ be zero on the exit boundary and $\mathbf{n}\cdot\nabla \varphi_j$ be zero on the reflecting boundary. Therefore, the adjoint operator $\mathcal{L}^*$
of $\mathcal{L}$ is the differential operator whose domain consists of continuous functions $\varphi=(\varphi_j)$ on $\mathcal{R}$, whose (weak) derivatives up  to second order are square integrable, having value $0$ on the exit boundary and normal derivative $0$ on the reflecting boundary. On these functions,
$$\left(\mathcal{L}^*\varphi\right)_j= \nabla\cdot(D_j\nabla\varphi_i)+\mathbf{v}_j\cdot \nabla\varphi_j+\sum_k\kappa_{jk}\varphi_k.$$

Notation: When applying $\mathcal{A}$ or $\mathcal{L}$ to $\rho(y,t|x,0)$ in the below calculations, we use a subscript such as in  $\mathcal{L}_x$ or $\mathcal{A}_y$ to indicate which variable it is being acted on.

We claim that $\mathcal{A}=\mathcal{L}^*$. This is seen as follows:
\begin{align*}
(\mathcal{A}\varphi)(x)&=\lim_{t\rightarrow 0}\frac1t\left(\int_{\mathcal{R}}\varphi(y)\rho(y,t|x,0)\, dy -\varphi(x)\right)\\[6pt]
&=\lim_{\epsilon\rightarrow 0}\lim_{t\rightarrow 0}\int_{\mathcal{R}}\varphi(y)\frac{\rho(y,t+\epsilon|x,0)-\rho(y,\epsilon|x,0)}{t}\, dy\\[6pt]
&=\lim_{\epsilon\rightarrow 0}\int_{\mathcal{R}}\varphi(y)\frac{\partial\rho}{\partial t}(y,\epsilon|x,0)\, dy\\[6pt]
&=\lim_{\epsilon\rightarrow 0}\int_{\mathcal{R}}\varphi(y)(\mathcal{L}_y\rho)(y,\epsilon|x,0)\, dy\\[6pt]
&=\lim_{\epsilon\rightarrow 0}\int_{\mathcal{R}}(\mathcal{L}^*\varphi)(y)\rho(y,\epsilon|x,0)\, dy\\[6pt]
&=\left(\mathcal{L}^*\varphi\right)(x).
\end{align*}
A similar argument to the above integration by parts computation also shows that $\mathcal{A}^*=\mathcal{L}$.  Since $\mathcal{A}$ is the relevant operator for the output composition problem, from this point on we dispense with the $\mathcal{L}, \mathcal{L}^*$ notation and only use $\mathcal{A}, \mathcal{A}^*$. We summarize below the definitions of $\mathcal{A}$ and $\mathcal{A}^*$. They are both defined on continuous vector-valued functions on $\mathcal{R}$ whose weak second derivatives are square integrable and satisfy:
\begin{align*}
(\mathcal{A}\varphi)_j  &= \nabla\cdot(D_j\nabla\varphi_j)+\mathbf{v}_j\cdot \nabla\varphi_j+\sum_k\kappa_{jk}\varphi_k\\[6pt]
\varphi_j &=0 \text{ on } \partial_{\text{\tiny exit}}\mathcal{R}\\[6pt]
\mathbf{n}\cdot\nabla\varphi_j &=0 \text{ on } \partial_{\text{\tiny reflect}}\mathcal{R}
\end{align*}
and 
\begin{align*}
(\mathcal{A}^*\psi)_j & = \nabla\cdot(D_j\nabla\psi_j-\psi_j\mathbf{v}_j)+ \sum_k\psi_k\kappa_{kj}\\[6pt]
\psi_j &=0 \text{ on } \partial_{\text{\tiny exit}}\mathcal{R}\\[6pt]
\mathbf{n}\cdot(\psi_j\mathbf{v}_j -D_j\nabla\psi_j)  &=0 \text{ on } \partial_{\text{\tiny reflect}}\mathcal{R}.
\end{align*}
As already noted, $\mathcal{A}$ generates a one-parameter semigroup $P_t$ such that 
$$(P_t\varphi)(x) = \int_{\mathcal{R}} \varphi(y)\rho(y,t|x,0)\, dy. $$
The adjoint semigroup is $P^*_t$ is characterized by   $\langle P^*_t\psi,\varphi\rangle=\langle \psi, P_t\varphi\rangle$. It is also an integral operator with (matrix) kernel denoted $\rho^*_{ij}(x,t|y,s)$. We 
  summarize here the defining  properties of these two integral kernels:
 \begin{itemize}
 \item $\rho(x,t|y,s)$, $t>s$,  satisfies
 $$ \frac{\partial \rho_{ij}}{\partial t} = \nabla\cdot(D_i \nabla\rho_{ij} - \rho_{ij}\mathbf{v}_i)+\sum_{k}\rho_{kj}\kappa_{ki} =(\mathcal{A}^*\rho_{\cdot j})_i$$
 where the derivatives are in $x$. Further, the boundary conditions
 $\rho_{ij}(x,t|y,s)=0$
 for $x$ on the exit boundary and  $\mathbf{n}\cdot(D_i\nabla\rho_{ij}-\rho_{ij}\mathbf{v}_i)=0$ for $x$ on the reflecting boundary
 and the initial condition
 $$\int_{\mathcal{R}}\rho(x,t|y,s)\varphi(y)\, dy\rightarrow \varphi(x) \text{ as } t\downarrow s$$
 hold for any given $\varphi$. 
 \item
  $\rho^*(x,t|y,s)$, $t<s$,  satisfies
 $$ -\frac{\partial \rho^*_{ij}}{\partial t} = \nabla\cdot(D_i \nabla\rho^*_{ij}) + \mathbf{v}_i\cdot\nabla \rho^*_{ij}+\sum_{k}\kappa_{ik}\rho^*_{kj} =(\mathcal{A}\rho^*_{\cdot j})_i$$
 where the derivatives are in $x$. Further, the boundary conditions
 $\rho^*_{ij}(x,t|y,s)=0$
 for $x$ on the exit boundary and  $\mathbf{n}\cdot\nabla\rho^*_{ij}=0$ for $x$ on the reflecting boundary
 and the initial condition
 $$\int_{\mathcal{R}}\rho^*(x,t|y,s)\varphi(y)\, dy\rightarrow \varphi(x) \text{ as } t\uparrow s$$
 hold for any given $\varphi$. 
 \end{itemize} 
  It can be further verified using a relatively standard argument for systems of parabolic partial differential equations (see \cite{Friedman}) that
  $$ \rho^*(y,s|x,t)=\rho(x,t|y,s)^\intercal$$
where $\intercal$ indicates matrix transpose. Finally, as the coefficients $D_i, \mathbf{v}_i, \kappa_{ij}$ do not depend on time $t$ explicitly, we have
$$\rho(x,t|y,s)=\rho(x,t-s|y,0). $$

\subsection{The output composition matrix}
The main quantity of interest in this work is the total amount of species $i$ which is produced after the reactor is fully evacuated. Let us represent this quantity by $A_i$. In analytic form, 
$$A_i = \lim_{T\rightarrow \infty}\int_0^T \int_{\partial_{\text{\tiny exit}}\mathcal{R}} \mathbf{n}(x)\cdot\mathbf{j}_i(x,t)\, dA(x)\, dt. $$
Here ${\partial_\text{\tiny exit}}\mathcal{R}$ indicates the exit boundary of the reactor and $dA(x)$ is the element of surface area. (The volume element will be written simply $dx$.) The integral over the exit boundary
gives the rate of flow of species $i$ out of the reactor and its integral over the time interval $[0,T]$ is the amount of $i$ that escapes by time $T$. An application of the divergence theorem (using that $\mathbf{n}(x)\cdot \mathbf{j}_i(x,t)=0$ on the reflecting boundary of $\mathcal{R}$) together with the reaction-transport equation yields
$$ A_i=\lim_{T\rightarrow \infty}  \int_0^T \int_{\mathcal{R}} -\frac{\partial c_i}{\partial t}(x,t)\, dx\, dt +\sum_{j}\int_0^\infty \int_{\mathcal{R}} c_j(x,t)\kappa_{ji}(x)\, dx\, dt.$$
Note that 
$$ \int_0^T \int_{\mathcal{R}} \frac{\partial c_i}{\partial t}(x,t)\, dx\, dt =\int_0^T \frac{d}{dt}\int_{\mathcal{R}} c_i(x,t)\, dx\, dt = Q_i(T)-Q_i(0), $$
where $Q_i(t):=\int_{\mathcal{R}} c_i(x,t)\, dx$ is the quantity of $i$ still left in the reactor by time $t$. Since in the long run the reactor is fully evacuated, 
\begin{equation}\label{Ai}
A_i = Q_i(0) +\sum_j \int_0^\infty \int_{\mathcal{R}} c_j(x,t)\kappa_{ji}(x)\, dx\, dt.
\end{equation}

Let $A_{ij}(y)$ indicate the amount (molar fraction) of species $i$ produced by the system given that, at the initial time, a unit pulse of $j$ is injected into the reactor at position $y$. This quantity can be expressed using the fundamental solution as
$$A_{ij}(y)=\delta_{ij} +\sum_k\int_0^\infty  \left[\int_{\mathcal{R}}\rho_{kj}(x,t|y,0)\kappa_{ki}(x)\, dx\right]\, dt . $$
 In matrix form, and slightly simplifying the notation $\rho(x,t|y):=\rho(x,t|y,0)$,
\begin{align*}
A(y)&=I + \int_0^\infty \int_{\mathcal{R}} \kappa^\intercal(x)\rho(x,t|y)\, dx\, dt\\[6pt]
&= I +\int_0^\infty \int_{\mathcal{R}}\kappa^\intercal(x)\rho^*(y,-t|x)^\intercal\, dx\, dt.
\end{align*}
Here $I$ is  the identity $N\times N$-matrix. The transpose of $A(y)$ is
$$ A^\intercal(y)=I+\int_0^\infty\int_{\mathcal{R}}\rho^*(y,-t|x)\kappa(x)\, dx\, dt.$$
Note that $\mathcal{A}I=\kappa(y)$ and that
$\frac{\partial}{\partial t}\rho^*(y,-t|x) =\mathcal{A}\rho^*(y,-t|x).$ 

We can now state the boundary-value problem satisfied by $A^\intercal(y)$. 
\begin{align*}
(\mathcal{A}A^\intercal)(y)&=\kappa(y) +\int_0^\infty \int_{\mathcal{R}}\mathcal{A}_y\rho^*(y,-t|x)\, \kappa(x)\, dx\, dt\\[6pt]
&=\kappa(y) +\int_0^\infty \frac{d}{dt}\int_\mathcal{R} \rho^*(y,-t|x)\kappa(x)\, dx\, dt\\[6pt]
&=\kappa(y) +\lim_{\epsilon\rightarrow 0}\lim_{T\rightarrow \infty} \int_{\mathcal{R}}\left[\rho^*(y,-T|x)-\rho^*(y,-\epsilon|x)\right]\kappa(x)\, dx.
\end{align*}
Now observe that $\rho^*(y,-T|x)= \rho(x,T|y,0)^\intercal$ will tend towards $0$ as $T\rightarrow \infty$. This is because, in the long run, the open reactor will 
be fully emptied. Furthermore $\rho^*(y,-\epsilon|x)\rightarrow \delta(x-y)I$ as $\epsilon\rightarrow 0$. 
We conclude that $(\mathcal{A}A^\intercal)(y)=\kappa(y)-\kappa(y)=0.$

When $y$ approaches the exit boundary, $\rho(x,t|y)$ approaches $0$, hence $A_{ij}(y)\rightarrow\delta_{ij}$. It also follows from the general properties of $\rho$ and
$\rho^*$ that $A^\intercal$ satisfies the Neumann condition on the reflecting boundary. Thus we obtain the following central result. (We now use $x$ for the position variable in $A$.)
\begin{conclusion}[Boundary-value problem for the output composition matrix]
The transpose $f(x)=A^\intercal(x)$ of the matrix-valued output composition  function $A(x)$ satisfies the equation $\mathcal{A}f=0$ on $\mathcal{R}$ together with
the Neumann condition on the reflecting boundary of $\mathcal{R}$ and $f(x)=I$ on  exit boundary points. 
\end{conclusion}

\section{Reduction to network reactors}\label{reduction}
We wish now to rewrite the boundary-value problem for the solution to the OCP so that it applies to network reactors. The main assumptions are that 
the transport coefficients $D_i$ and $\mathbf{v}_i$ are constant on cross-sections of reactor pipes (see Figure \ref{Juncture}) and that the output composition matrix $f(\mathbf{x})=A^\intercal(\mathbf{x})$ is well-approximated by functions  that are  constant on those cross-sections.  Due to the Neumann condition on the reflecting boundary, this assumption can be expected to hold well if pipes are long and narrow. Recalling that the reaction coefficients are zero outside junctures, 
then the partial differential operator $\mathcal{A}$ reduces to
$$\mathcal{A}\varphi_i =\nabla\cdot(D_i\nabla\varphi_i)+\mathbf{v}_i\cdot\nabla\varphi_i=\frac{d}{dx}\left(D_i^b\frac{d\varphi_i}{dx}\right) +\nu_i^b\frac{d\varphi_i}{dx}=0, $$
where $x$ is arc-length parameter along the branch $b$ (the center axis of the pipe) and a choice of orientation of $b$ has been made so that the sign of $\nu_i^b$ is defined. To these equations (indexed by $i$ and $b$) we need to add conditions on nodes, obtained as junctures are imagined to shrink to points. 
It is natural to suppose that, in the limit,  $f$ is still  continuous at nodes (its value  at a node coincides with the values of the limits as the node is approached from the direction of any of the branches attached to it.)  We further need to specify conditions on the derivatives of $f$ along branch directions at a node. This is done next.

\subsection{Node conditions}\label{node conditions}
The conditions on $f$ and its first directional  derivatives at a node $n$ can be determined by integrating the (matrix) equation
$ \mathcal{A}f=0$ over the   juncture indexed by $n$, applying the divergence theorem, and recalling the relationship between $\kappa$ and $K$ described in 
Subsection \ref{kappaK}. (See the notation described in Figure \ref{Juncture}.) Let us write $\varphi_i=f_{ij}$ for a fixed $j$. Thus we integrate term-by-term  the equation
$$\nabla\cdot(D_i\nabla \varphi_i) + \mathbf{v}_i\cdot \nabla\varphi_i +\sum_k \kappa_{ik} \varphi_k=0$$
on the juncture $\mathcal{R}_\epsilon(n)$. Since $\varphi_i$ has zero normal derivative at the reflecting boundary of $\mathcal{R}$, 
$$\int_{\mathcal{R}_\epsilon(n)}\nabla\cdot(D_i\nabla \varphi_i)\, dV = \sum_b \int_{\mathcal{A}_\epsilon(n,b)} D_i\nabla\varphi_i \cdot \mathbf{u}_{b}(n)\, dA\approx \sum_b A_\epsilon(n,b)D^b_i(n) \frac{d\varphi^b_i}{dx}(n).$$
Here $\mathcal{A}_\epsilon(n,b)$ is the disc component of the boundary of $\mathcal{R}_\epsilon(n)$ that attaches to the pipe indicated by $b$ and, as already defined, $A_\epsilon(n,b)$ is the area of this disc. We assume that the parametrization is such that $x=0$ corresponds to $n$.  $V$ and $A$ are the volume and area elements in integration. 

The volume integral of $\mathbf{v}_i\cdot \nabla\varphi_i$ is proportional to the volume $V_\epsilon(n)$ of the juncture. Let us write it as 
$C_\epsilon(n)V_\epsilon(n)$, where $C_\epsilon(n)$ is   the average  value of the integrand on $\mathcal{R}_\epsilon(n)$. 
The integral of $\kappa_k\varphi_{ik}$ can be approximated using the characterization of $K_{ik}(n)$ (see Section \ref{kappaK}):
$$ \int_{\mathcal{R}_\epsilon(n)}\kappa_{ik}(\mathbf{x})\varphi_k(\mathbf{x})\, dV\approx A_\epsilon(n)K_{ik}(n)\varphi_k(n).$$
Here we recall that $A_\epsilon(n)$ is the sum of the $A_\epsilon(n,b)$ over the branches $b$ attached to $n$.  Adding these three terms and dividing by $A_\epsilon(n)$ (recall that $p(n,b)=A_\epsilon(n,b)/A_\epsilon(n)$), we obtain, to first order in $\epsilon$,
$$  \sum_b p(n,b)D^b_i(n) \frac{d\varphi^b_i}{dx}(n) + C_\epsilon(n) \frac{V_\epsilon(n)}{A_\epsilon(n)} + \sum_k  K_{ik}(n)\varphi_k(n)=0.$$
Now recall that $\epsilon A_\epsilon(n)/V_\epsilon(n)$ converges to a dimensionless number $\chi(n)$ characteristic of the juncture geometry. Thus
$ {V_\epsilon(n)}/{A_\epsilon(n)}$ is of the order $\epsilon$. Eliminating this term and passing to the limit as $\epsilon$ approaches $0$ yields
\begin{equation}\label{node_condition}
 \sum_b p(n,b)D^b_i(n) \frac{d\varphi^b_i}{dx}(n)+ \sum_k  K_{ik}(n)\varphi_k(n)=0.
\end{equation}
\subsection{Invertibility of the matrix $\Lambda$}\label{invertibility}
Recall the definition of the matrix $\Lambda$ in Equations (\ref{Lambda_def}) or (\ref{matrix}). We wish to provide here sufficient conditions for $\Lambda$ to be invertible based on the form of this matrix,  independently of the boundary-value problem from which  it originated. Let $\lambda_{ij}$ be the $(i,j)$-matrix element of $\Lambda$.  We
first note the following two properties of these elements:
\begin{enumerate}
\item\label{dia_dom} For each row of $\Lambda$ with row index $i$ we have 
 $|\lambda_{ii}|\geq \sum_{j:j\neq i} |\lambda_{ij}|.$
\item\label{dia_dom2} For at least one $i$ (in fact, for   those $i$ for which $\eta(n_i,n_{\text{\tiny exit}})>0$), $|\lambda_{ii}|> \sum_{j:j\neq i} |\lambda_{ij}|.$
\end{enumerate}
In checking property \ref{dia_dom}, notice that  $\sum_j\eta(n_i,n_j)=1$ and $\sum_{s}\tilde{K}_{rs}(n_i)=0.$ 

We now make the assumption that the network reactor is connected in the following sense: if we remove the exit nodes as well as all the branches attached to those nodes, then the resulting network does 
not disconnect. There is  no loss of generality in making this assumption since, otherwise, the output composition matrix would only depend on the connected piece that received the initial injection of reactants.

Define $\tilde{K}=\sum_{n}\tilde{K}(n)$, where the sum is over   the active nodes of the network reactor. Invertibility of $\Lambda$ can be    established by general facts about matrices   if we make the    further assumption that any substance among the $N$ species can be transformed into any other by a sequence of reactions among those encoded in $\tilde{K}$.  This assumption on $\tilde{K}$ together with  connectivity of the network imply that $\Lambda$ is an irreducible matrix in the sense of definition 6.2.25 of \cite{HJ}. Irreducibility plus the above properties \ref{dia_dom} and \ref{dia_dom2} are the conditions needed to apply  Corollary 6.2.27 in \cite{HJ}, which implies the desired invertibility result.

\section{Conclusions}\label{conclusions}
In this paper we address the following basic problem for reaction-transport systems in open reactors, which we call the {\em Output Composition Problem} (OCP). We suppose that a    reactor in which   a system of reactions with linear rate functions can take place involving a number of gas species and solid catalysts. Reactor exits are kept at vacuum conditions. 
Reactions are assumed to occur at localized regions that are chemically active. These regions are connected to each other by narrow pipes so that the whole arrangement may be regarded as a network of small reactor units. We call such  configuration a {\em network-like reactor}. 
The OCP is then the problem of determining the composition of (i.e., the fractions of each participating substance in)  the collected gas after the reactor is fully emptied. 

We provide a solution to this problem in the form of a boundary-value problem for a system of time-independent partial differential equations, which provides the output composition more efficiently than   a direct approach based on solving    the reaction-advection-diffusion (time dependent) equations. This result holds for very general reactor domains.

By approximating network-like reactors with actual {\em network reactors}, which are networks of curves (branches) connected at points (nodes) with active regions reduced to nodes and transport coefficients (diffusivities and advection velocities) constant along branches,  the   boundary-value problem solving the OCP reduces to a system of linear algebraic equations that are easily solved by elementary means. 

Finally, we explain our method with several  representative examples and draw a number of general observations from them. The mathematical theory behind the approach is also given.

Moving forward,  we plan to apply    the methods developed in this paper to the description and interpretation of TAP data. We believe that such an analysis will provide an effective method for probing the details of compex catalytic mechanisms. 

\section{Glossary of main symbols}\label{symbols}
{\footnotesize $\begin{array}{lll}
\text{\bf Symbol} & \text{\bf description} & \text{\bf defined in section} \\[6pt]
\hline \\
\mathcal{A}_\epsilon(n,b), A_\epsilon(n,b)  & \text{coss-section of $b$ and its area at juncture $n$} & \ref{kappaK}  \\[6pt] 
\mathcal{A}=\mathcal{L}^*& \text{differential operator $\mathcal{A}$, the adjoint of $\mathcal{L}$} & \ref{proof_1} \\[6pt]
b, \mathbf{b}=(n,n'), \overline{\mathbf{b}}& \text{branch, oriented branche, reverse orientation} & \ref{def_notat} \\[6pt]
c_i(\mathbf{x},t) & \text{concentration of $i$ at $\mathbf{x}$ in $\mathcal{R}$ at time $t$} & \ref{Sub_bvp} \\[6pt]
D_i(\mathbf{x}) & \text{diffusion coefficient of $i$ at $\mathbf{x}$ in $\mathcal{R}$} & \ref{Sub_bvp} \\[6pt]
\text{deg}(n) & \text{degree of node $n$} & \ref{def_notat} \\[6pt]
f_{ij}(n) & \text{output composition matrix} &  \ref{Sub_bvp} \\[6pt]
\mathbf{j}_i(\mathbf{x}) & \text{flux density of $i$ at $\mathbf{x}$} & \ref{Sub_bvp} \\[6pt]
\kappa_{ij}(\mathbf{x}), K_{ij}(n), \tilde{K}_{ij}(n) & \text{reaction coefficients in reaction domain and on network} &  \ref{def_notat}, \ref{kappaK}, \ref{solution}\\[6pt]
\ell(b), \tilde{\ell}(\mathbf{b}) & \text{length of branch, velocity-adjusted length} & \ref{def_notat}, \ref{vel_adjusted} \\[6pt]
\mathcal{L} & \text{differential operator of reaction-transport equation} & \ref{proof_1} \\[6pt]
n & \text{network node} & \ref{def_notat} \\[6pt]
\mathbf{n}(\mathbf{x}) & \text{normal vector field to boundary of $\mathcal{R}$} & \ref{Sub_bvp} \\[6pt]
p(n,b) & \text{cross-section area fraction of $b$ at juncture $n$} & \ref{def_notat}, \ref{kappaK} \\[6pt]
\mathcal{R} & \text{network-like reactor domain, network reactor}  & \ref{Sub_bvp} \\[6pt]
\mathbf{v}_i(\mathbf{x}), \nu_i(\mathbf{b}), \nu_i^b & \text{advection velocity of $i$} & \ref{Sub_bvp}, \ref{def_notat} \\[6pt]
\eta_i(n,b) & \text{coefficients of matrix $\Lambda$} & \ref{solution}, \ref{examples}\\[6pt]
\varphi_{\mathbf{b}}(x) & \text{arc-length parametrization of $\mathbf{b}$} & \ref{def_notat} \\[6pt]
\Lambda(n_i,n_j), \lambda  & \text{coefficient matrix and vector in linear algebraic equations for $f$} & \ref{solution} \\[6pt]
\xi_i(n,b) & \text{branch coefficients in definition of $\eta_i(n,b)$} & \ref{solution}, \ref{examples} \\[6pt]
\rho(x,t|y,s) & \text{fundamental solution of reaction-transport equation} & \ref{proof_1} \\[6pt]
\end{array}$
}

\bibliographystyle{siam}

\end{document}